\documentclass[aps,prl,reprint,groupedaddress,superscriptaddress]{revtex4-2}
\usepackage{lineno,hyperref}
\usepackage{xcolor}
\usepackage[separate-uncertainty=true,multi-part-units=single]{siunitx}
\usepackage{graphicx}
\usepackage{amsmath}
\usepackage{cleveref} 
\usepackage{breqn}
\usepackage{soul}

\crefname{equation}{Eqn.}{Eq.}
\Crefname{equation}{Equation}{Equations}
\crefname{figure}{Fig.}{Figs.}
\Crefname{figure}{Figure}{Figures.}
\crefname{section}{Sec.}{Secs.}
\crefname{table}{Table S}{Tables S}

\renewcommand*{\vec}[1]{\boldsymbol{#1}}

\makeatletter
\let\cat@comma@active\@empty
\makeatother

\begin{document}


\title{Probing lipid membrane bending mechanics using gold nanorod  tracking}


\author{Mehdi Molaei} 
\affiliation{Chemical and Biomolecular Engineering, University of Pennsylvania, Philadelphia, PA 19104}

\author{Sreeja Kutti Kandy} 
\affiliation{Bioengineering, University of Pennsylvania, Philadelphia, PA 19104}

\author{Zachary T. Graber} 
\affiliation{Chemistry, University of Pennsylvania, Philadelphia, PA 19104}

\author{Tobias Baumgart}
\affiliation{Chemistry, University of Pennsylvania, Philadelphia, PA 19104}

\author{Ravi Radhakrishnan}
\affiliation{Bioengineering, University of Pennsylvania, Philadelphia, PA 19104}

\author{John C. Crocker}
\thanks{Corresponding author. jcrocker@seas.upenn.edu}
\affiliation{Chemical and Biomolecular Engineering, University of Pennsylvania, Philadelphia, PA 19104}




\begin{abstract}

Lipid bilayer membranes undergo rapid bending undulations with wavelengths from tens of nanometers to tens of microns due to thermal fluctuations.  Here, we probe such undulations and  the membranes' mechanics by measuring the time-varying orientation of single gold nanorods (GNRs) adhered to the membrane, using high-speed dark field microscopy. In a lipid vesicle, such measurements allow the determination of the membrane's viscosity, bending rigidity and tension as well as the friction coefficient for sliding of the monolayers over one another. The in-plane rotation of the GNR is hindered by undulations in a membrane tension dependent manner, consistent with simulations. The motion of single GNRs adhered to the plasma membrane of living cultured cells similarly reveals that membrane's complex physics and coupling to the cell's actomyosin cortex.  

\end{abstract}



\maketitle

The plasma membrane of cells, beyond serving as a boundary between intracellular and extracellular spaces, mediates force transmission, and information and material flow between inside and outside, via cytoskeletal adhesion proteins, signal receptor proteins and endocytic and exocytic vesicle formation.
This fluidic lipid membrane displays time-dependent local curvature or undulations, which are controlled by its bending modulus, excess area and membrane tension.
The complex physics of model lipid membranes, as in giant unilamellar vesicles (GUV) formed of purified lipids, is relatively well understood, having been studied by a variety of techniques \cite{graber_cations_2017, shi_cell_2018, fournier_fluctuation_2004, dimova_recent_2014,kilpatrick_nanomechanics_2015, kocun_pulling_2012, engelhardt_viscoelastic_1984,dimova_vesicles_2009,kummrow_deformation_1991}.
In contrast, the physical properties and dynamics of the plasma membrane of cells are notoriously difficult to measure \cite{sheetz_modulation_1996, dai_mechanical_1995, capraro_curvature_2010, hochmuth_micropipette_2000, park_measurement_2010, feng_association_2021}.
Much of this difficulty stems from the membrane being attached to the underlying viscoelastic actin cortex by adhesion proteins, having a mean spacing of a few hundred nanometers—--deformations and  undulations on longer lengthscales are simply those of the composite membrane-cortex structure.  Nevertheless, it is the undulations of the plasma membrane itself on the \SI{100}{\nano\meter} lengthscale that couple to vesicle trafficking, membrane curvature and tension sensing as well as membrane protein binding and assembly. 

Here we utilize orientational tracking of single gold nanorods (GNRs) to characterize the small lengthscale undulations and nanorheology of reconstituted lipid GUVs and the plasma membrane of cultured cells.
While the in-plane rotational motion of anisotropic probes have previously been applied to infer membrane viscosity \cite{hormel_measuring_2014}, we exploit precise three-dimensional orientational tracking to access the dynamics of lipid membrane bending undulations.
By developing a model for out of plane angular motion of GNRs on GUVs, we determine physicochemical properties of the lipid bilayers including the bending modulus, intermonolayer friction coefficient and membrane tension.
The detailed motion of a nanorod on an undulatory membrane is also studied through Monte Carlo simulation.
The angular motion of GNRs bound to the plasma membrane of a Huh7 cell is tracked to investigate dynamics of a fluctuating plasma membrane.

We first apply our method to study the membrane dynamics of GUVs with different lipid compositions and at different tensions.
Experimentally, we attach GNRs (with nominal length \(l=\SI{140}{\nano\meter}\) and diameter \(2a=\SI{40}{\nano\meter}\)) to GUVs having either 1:1 composition of 1,2 dioleoyl-sn-glycero-3-phosphocholine (DOPC) and 1,2-dioleoyl-sn-glycero-3-phospho-L-serine (DOPS), or 1:1:1 composition of DOPS-DOPC-Cholesterol, using magainin 2 peptides.
We also change the effective surface tension of the cholesterol GUVs by changing osmolarity of the buffer solution and form both tense and floppy GUVs, see Supplemental Material (SM) for details \cite{noauthor_see_nodate-1}\nocite{xie_nuclear_2009, angelova_liposome_1986,  tian_sorting_2009, khan_random_2014}.
We image the GNRs using a custom-built laser-illuminated dark field microscope \cite{molaei_nanoscale_2018}. We track the centroid, \(x\) and \(y\), of the GNRs over time \(t\) by particle tracking \cite{crocker_methods_1996, rose_particle_2020}.
Polarimetric analysis of the scattered light from single nanorods allows the reconstruction of the orientation vector \(\hat{\vec{u}}(t)\) of their major axis in 3D space with a precision of about \(\SI{1}{\degree}\), at several thousand measurements per second.
The center coordinates, \(x_c,y_c\), and  radius, \(R\), of the GUVs are determined by phase contrast microscopy. 

\begin{figure}
 	\centering
		 \includegraphics[width=0.49\textwidth]{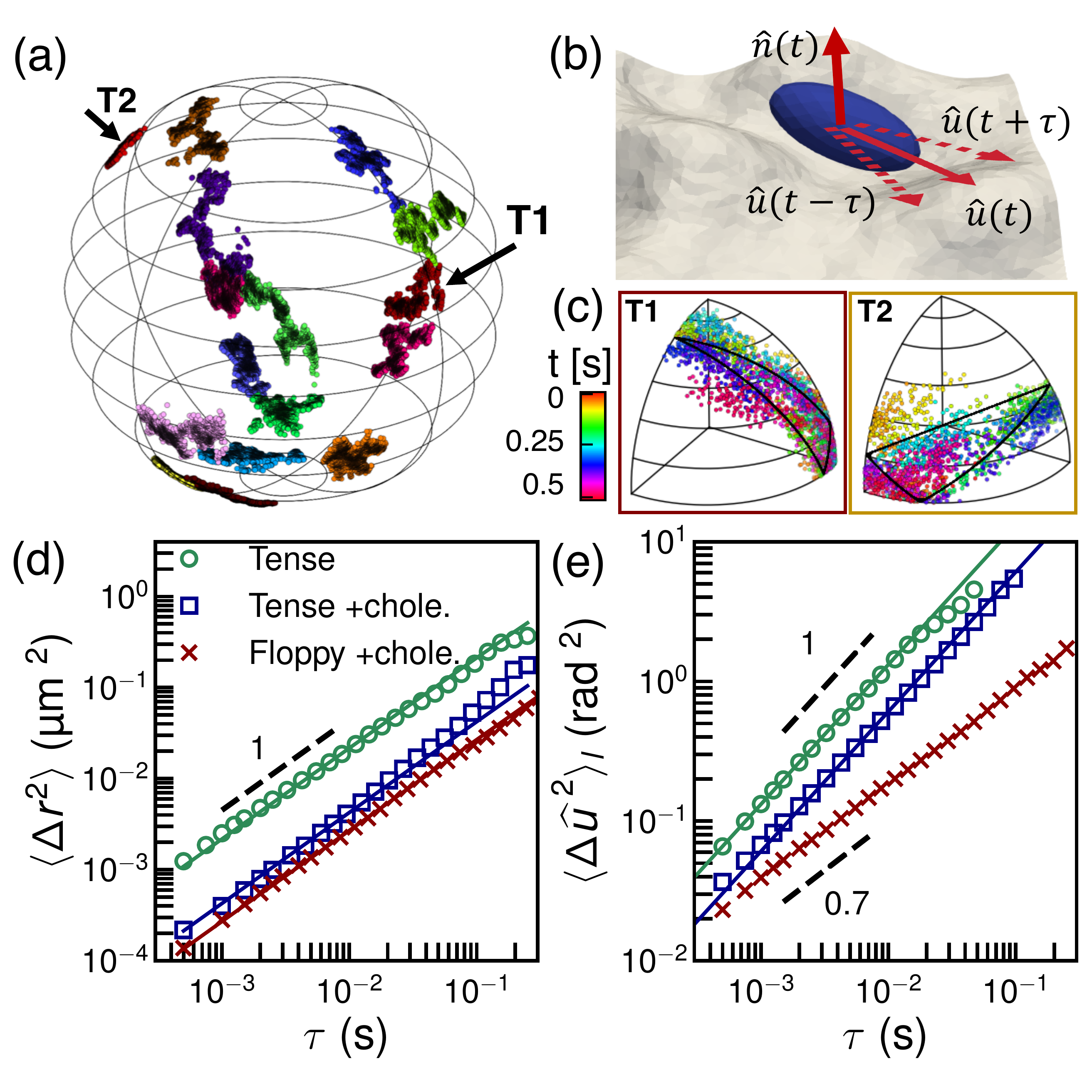}
 			\caption{Thermal motion of GNRs on GUVs, varying membrane tension and viscosity. (a) Reconstructed 3D trajectories of the GNR on different parts of a tense (isoosmolar) GUV, each lasting \(\approx \SI{1}{\second}\); the GUV radius is \(R=\SI{7}{\micro\meter}\). (b) Schematic of a membrane bond nanoprobe, its orientation vector \(\hat{\vec{u}}\) and the membrane normal vector \(\hat{\vec{n}}\).  (c) Two typical angular trajectories of GNR orientation, \(\hat{\vec{u}}\), corresponding to the labels T1 and T2 in panel (a), showing rapid in-plane rotation mapped into one octant; loops shown by black lines indicate projection of planes with \(\hat{\vec{n}}=\langle \hat{\vec{r}}(t) \rangle_t\) into the same octant. (d) MSDs of the GNRs show diffusive translational motion, open symbols, and fits to \(4D_t\tau\), solid lines. (e)  MSADs of the GNRs on the tense GUVs display diffusive motion, on the floppy GUV subdiffusive motion, open symbols, fits to \(4D_r\tau\) solid lines.}
	 \label{MSD}
 \end{figure}

The experimental 3D trajectories of single GNRs over time \(t\), \(\vec{r}(t)=[x(t)-x_c,~ y(t)-y_c,~ z(t)]\) diffusing at different locations on the no-cholesterol GUV is shown in \cref{MSD}(a), where \(z=[R^2-(x-x_c)^2-(y-y_c)^2]^{1/2}\) is the depth of the GNR. The random motions in all three dimensions are Gaussian distributed \cite[Fig. S1]{noauthor_see_nodate-1}). The coordinates of the nanorod orientation and membrane normal are shown in \cref{MSD}(b), and typical angular trajectories are shown in \cref{MSD}(c). The translational mean-squared displacement (MSD) of GNRs, \(\langle\Delta \vec{r}^2(\tau)\rangle = \langle\left(\vec{r}(t+\tau)-\vec{r}(t)\right)^2\rangle_t\) on different GUVs are shown in \cref{MSD}(d), where \(\langle\cdot\rangle_t\) denotes a time average. The MSD of a particle diffusing over the surface of a sphere is bounded at long times \cite{noauthor_see_nodate-1}, but satisfies \(\langle\Delta \vec{r}^2 (\tau)\rangle=4D_t\tau\) for \(\tau\ll R^2/D_t\), where \(D_t=k_BT/\gamma_t\) is the translational diffusivity, and \(\gamma_t\), and \(k_BT\) are the translational drag coefficient and thermal energy.

The mean-squared angular displacement (MSAD), \(\langle\Delta \hat{\vec{u}}^2(\tau)\rangle=\langle(\hat{\vec{u}}(t+\tau)-\hat{\vec{u}}(t))^2\rangle_t\), of the GNRs is also measured to determine the in-plane rotational drag coefficient of the rods, \(\gamma_r\). With \(\hat{\vec{u}}\) projected to an octant of the full sphere by the symmetry of our measurement technique \cite{molaei_nanoscale_2018}, the MSAD is also bounded at long times\cite{noauthor_see_nodate-1}. Using a simple mapping method developed in a previous study \cite{molaei_nanoscale_2018}, the measured MSADs can be converted into unbounded ones, \(\langle\Delta \hat{\vec{u}}^2(\tau)\rangle_l=4/k\left[ -\ln{(1-\langle\Delta \hat{\vec{u}}^2(\tau)\rangle/\Delta \hat{\vec{u}}_\infty^2)}  \right]^{1/\xi}\),
where \(k=4.04\), \(\xi=0.85\), and \(\Delta \hat{\vec{u}}_\infty^2=2-16/\pi^2\) is the asymptote value \cite[Fig. S2(c)]{noauthor_see_nodate-1}. For the simple rotational diffusion case, this MSAD then satisfies
\(\langle\Delta\hat{\vec{u}}^2(\tau)\rangle_l=4D_r\tau\), where \(D_r=k_BT/\gamma_r\) is the rotational diffusivity, and \(\gamma_r\) is the rotational drag coefficient.

Fitting the measured MSDs and MSADs yields diffusivities \(D_t\) and \(D_r\) for single GNRs and their drag coefficients \(\gamma_t\) and \(\gamma_r\) on different GUVs. In general, these drag coefficients depend on the membrane viscosity, \(\eta_m\), the size of the nanorod, and the bulk fluid viscosity \(\eta\) \cite{levine_dynamics_2004, hormel_measuring_2014}. 
For the GNR on the tense GUV without cholesterol, we obtain \(D_t=\SI{0.52 +- 0.06}{\micro\meter^2\per\second}\) and \(D_r=\SI{27.7 +- 3.1}{rad^2\per\second}\).
Using a theoretical model for \(\gamma_t\) and \(\gamma_r\) \cite{saffman_brownian_1975, levine_dynamics_2004, naji_corrections_2007} and assuming a bulk fluid viscosity \(\eta=\SI{1}{\milli\pascal\second}\) (see SM  \cite{noauthor_see_nodate-1}), we estimate the membrane viscosity \(\eta_m=\SI{1.2 +-0.1}{\nano\pascal\second\meter}\), in good agreement with literature values \cite{stanich_coarsening_2013, hormel_measuring_2014, petrov_translational_2012, honerkamp-smith_membrane_2013}. This result also indicates that the Saffman-Delbr\"uk length scale, set by the ratio of the membrane and bulk viscosities, \(l_m=\eta_m/\eta\approx\SI{1}{\micro\meter}\), is an order of magnitude larger than the GNR length, ensuring that its motion is dominated by membrane mechanics rather than the bulk fluid \cite{hormel_measuring_2014}. 

The membrane viscosity of the tense GUV with cholesterol is estimated as \(\eta_m=\SI{17.7 +- 0.2}{\nano\pascal\second\meter}\) which is an order of magnitude larger than \(\eta_m\) of the no-cholesterol GUV. The effect of cholesterol on the fluidity of lipid membranes depends on  temperature \cite{kucerka_fluid_2011} and lipid composition \cite{espinosa_shear_2011}. While cholesterol increases the fluidity of lipid membranes with saturated or monounsaturated fatty acids tails \cite{espinosa_shear_2011}; it raises the viscosity of unsaturated tail lipids, such as DOPC-DOPS here, by about one order of magnitude \cite{bacia_sterol_2005}.
Interestingly, the MSAD of the GNR on the floppy GUV is lower than the same composition tense GUV and shows subdiffusive behavior, \cref{MSD}(e), which we argue below is due to transient caging by membrane undulations

While a GNR is randomly diffusing over a GUV, the membrane of the GUV also is undergoing thermal undulation causing tilting motion of the nanorods lying prone on the membrane (an orientation that maximizes their adhesion energy). 
We demonstrate here that the membrane normal direction \(\hat{\vec{n}}\) can be estimated by fitting the time dependent nanorod orientation over a few measurements to a plane, as illustrated in \cref{MSD}(b), using singular value decomposition \cite[Fig. S6(a)]{noauthor_see_nodate-1}).
This approach relies on the fact that in-plane rotational diffusion is typically much faster than the corresponding diffusion of the membrane normal, and is validated using Monte Carlo simulation, \cite[Fig. S6(b)]{noauthor_see_nodate-1}. 

\begin{figure}
    \centering
    \includegraphics[width=0.45\textwidth]{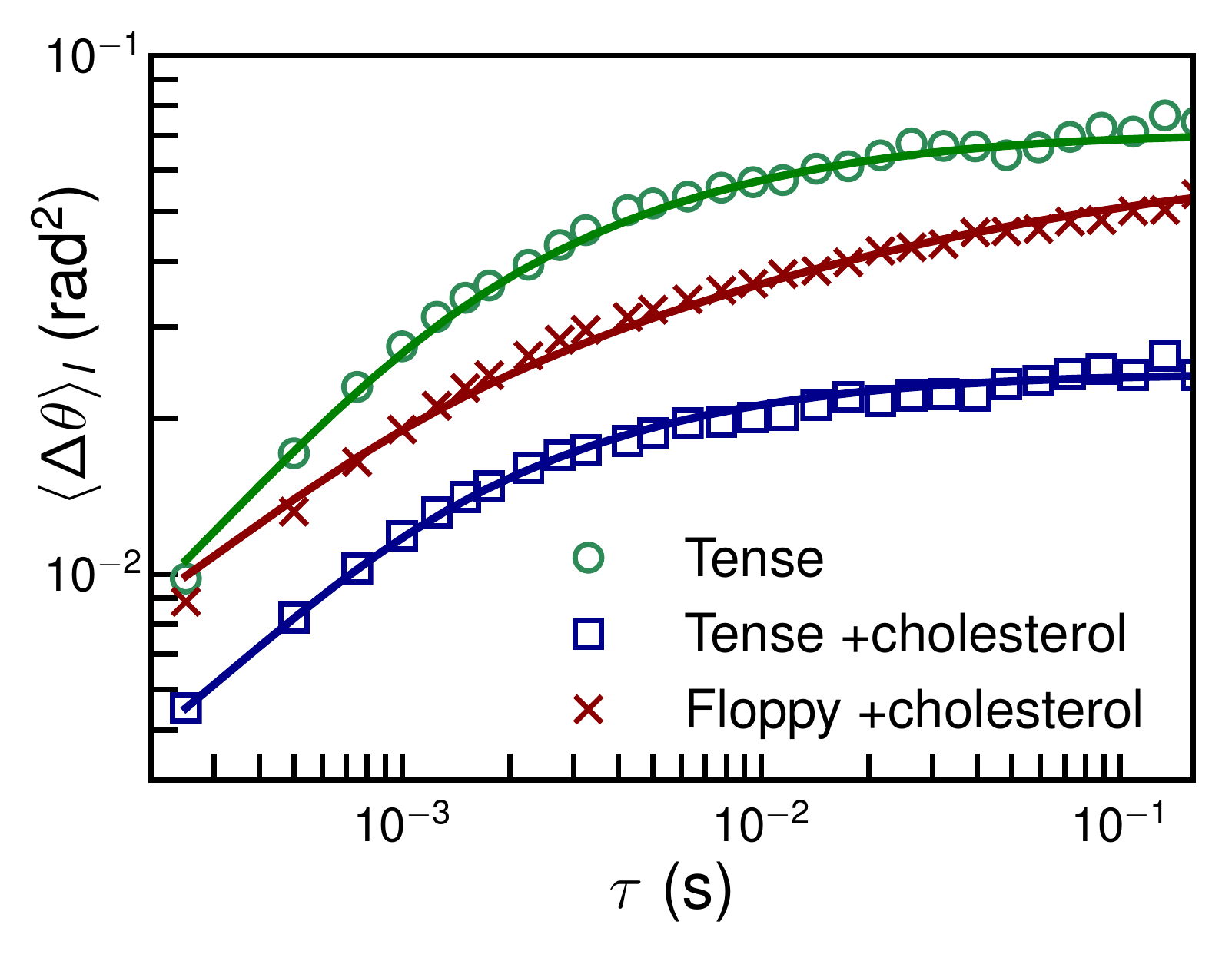}
    \caption{Out of plane angular motion of GNRs on different GUVs reveals the dynamic fluctuations of the membrane normal vector \(\hat{\vec{n}}\). Symbols are the measurement with solid lines fit to the model: Eqn. \ref{MSAD_normal_eq}.}
    \label{MSAD_normal}
\end{figure}

The dynamic fluctuations of the membrane normal can be characterized by the `out of plane' angle between the normal vector and the expected normal if the GUV were a perfect sphere: \(\theta=cos^{-1}\hat{\vec{n}}\cdot\hat{\vec{r}}\).
The MSDs of this out of plane motion \(\langle\Delta \theta^2(\tau)\rangle\), shown in \cref{MSAD_normal}, reveal dramatic differences in the undulation dynamics among different GUVs. 
Membrane undulatory dynamics has often been studied by analysing the bending fluctuations of the membranes, for example using optical microscopy. In the long-wavelength (small wavenumber) limit, in the Monge gauge, the spectrum of the height undulations of an elastic membrane varies as \( \langle h_q h_{-q} \rangle \sim 1/q^4 \). Since angular fluctuations correspond to the slope or derivative of the undulation height function, we expect them to vary as \( \langle \theta_q \theta_{-q} \rangle \sim 1/q^2 \) \cite{noauthor_see_nodate-1}. 
This suggests that angular measurements, as in our experiment, are far more sensitive to higher wavenumbers than direct height measurements, improving our measurement sensitivity to small amplitude, sub-micron wavelength undulations.   

Previous measurements of membrane fluctuations at higher wavenumber, such as small-angle X-ray scattering and neutron spin echo (NSE) spectroscopy \cite{boggara_effect_2010, nagao_probing_2017, yi_bending_2009, takeda_neutron_1999}, have revealed departures from the classical picture of a membrane as a thin fluid sheet coupled to a viscous bulk fluid \cite{brochard_frequency_1975}.  
At lengthscales closer to the membrane thickness, membrane viscosity and the friction of sliding between the two monolayers act as additional sources of dissipation and leads to a larger effective dynamic bending modulus than that observed in the long wavelength regime \cite{seifert_viscous_1993}. 

Undulations with different wavelengths relax at different rates, so the lag-time dependence of the normal fluctuations provides information about the amplitude of undulations at different wavelengths. Studies using MD simulation \cite{shkulipa_thermal_2006, watson_determining_2012, sadeghi_large-scale_2020} and high-speed video microscopy \cite{rodriguez-garcia_bimodal_2009} have shown that the correlation function of the membrane fluctuation relaxes with two q-dependent characteristic rates.  
The dynamics of long and short wavelengths undulations have different physical origins, and are separated by \(q_c \equiv 2\eta\varepsilon/b\tilde{\kappa}_c\), where \(\varepsilon\) is the compressibility modulus of the lipid bilayer and \(b\) is the friction coefficient between layers. \(\tilde{\kappa}_c=\kappa_c+2\varepsilon d^2\) is the renormalized bending rigidity accounting for the effect of the elastic stretching and compression, where \(\kappa_c\) is the bending rigidity, and \(d\) is the thickness of the lipid layer. 
At small wave number, \(q<q_c\), the slow relaxation mode is controlled by bending motion \(\omega_1=\kappa_c q^3/4\eta\), and the fast mode is driven by intermonolayer friction, \(\omega_2=\varepsilon q^2/4b\). For \(q>q_c\), however, the slow mode corresponds to the lateral density fluctuation of lipid molecules and intermonolayer friction, \(\omega_1=\frac{\varepsilon}{2b}\frac{\kappa_c}{\tilde{\kappa}_c}q^2\), and bulk viscosity quickly relaxes the bending motion with a rate of \(\omega_2=\frac{\tilde{\kappa}_c}{4\eta}q^3\). 
For this regime, adapting the Seifert-Langer (SL) model for height correlation function \cite{seifert_viscous_1993}, we develop a model for out of plane rotational motion of a GNR \cite{noauthor_see_nodate-1} 
\begin{dmath}
\label{MSAD_normal_eq}
    \langle\Delta\theta^2(\tau)\rangle=\frac{k_BT}{\pi}\int_{q_{min}}^{q_{max}}\frac{qdq}{\kappa_cq^2+\sigma}
    \left[1-\frac{\kappa_c}{\tilde{\kappa}_c}e^{-\omega_2\tau}-\left(1-\frac{\kappa_c}{\tilde{\kappa}_c}\right)e^{-\omega_1\tau}\right]+\langle\Delta\theta^2\rangle_{static},
\end{dmath}
where \(q_{min}=\pi/R\) and \(q_{max}=2\pi/l\) are the minimum and the maximum wave numbers set by the size of the GUVs and the GNRs respectively, \(\sigma\) is the surface tension, and \(\langle\Delta\theta^2\rangle_{static}\) is the measurement error. 
The asymptotic value of the out of plane displacement is controlled solely by bending rigidity and surface tension,
\begin{equation}\label{Eq:MSADinf}
\langle\Delta\theta^2_\infty\rangle=\frac{k_BT}{2\pi\kappa_c}\ln{\frac{\kappa_cq_{max}^2+\sigma}{\kappa_c q_{min}^2+\sigma}}.
\end{equation}
We use the variance of the measured \(\theta\) across multiple measurements to estimate  \(\langle\Delta\theta^2_\infty\rangle\), considering  \(\langle\Delta\theta^2(\tau)\rangle=2\langle\theta^2(t)\rangle_t-2\langle\theta(t)\theta(t+\tau)\rangle_t\) and the correlation function decays to zero at long time.  Thus, Eqn. \ref{Eq:MSADinf} provides a relationship between \(\kappa_c\) and \(\sigma\).

To model the out of plane motion of our nanorod data, as in \cref{MSAD_normal}, we fit Eqn. \ref{MSAD_normal_eq} to the measured \(\langle\Delta\theta^2(\tau)\rangle\) using \(\kappa_c\), \(b\), and \(\langle\Delta\theta^2\rangle_{static}\) as free parameters.  With this estimate of \(\kappa_c\), we compute the surface tension from Eqn. \ref{Eq:MSADinf}, and iteratively repeat the fitting with updated \(\sigma\) until the results converge. 
To limit the number of free parameters in the model, we choose nominal values from the literature for the compressibility modulus \(\varepsilon=\SI{0.1}{\newton\per\meter}\) and lipid layer thickness \(d=\SI{2}{\nano\meter}\) \cite{rodriguez-garcia_bimodal_2009}.

For the GUV with no-cholesterol, we find \(b=\SI{6.9 +- 0.5 e8}{\newton\second\per\meter^3}\), \(\kappa_c=\SI{7.3 +- 1.1}{}~k_BT\), and \(\sigma=\SI{1.1 +- 0.2e-6}{\newton\per\meter}\).
The standard error for \(\sigma\) is estimated from the standard error in \(\langle\theta^2(t)\rangle\) determined from more than 10 GNR trajectories with time spans of more than 2 minutes between each measurement. 
We confirmed the sensitivity of the measurement to the fitting parameters by performing a \(\chi^2\) test \cite[Fig. S8]{noauthor_see_nodate-1}. 
The values obtained from the fitting agree well with the ranges reported in the literature for \(b=\SI{0.5}{}-\SI{5e8}{\newton\second\per\meter^3}\) \cite{raphael_accelerated_1996, pott_dynamics_2002, rodriguez-garcia_bimodal_2009} and \(\kappa_c\approx\SI{10}{}~k_BT\) \cite{rodriguez-garcia_bimodal_2009, rawicz_effect_2000, evans_entropy-driven_1990, mell_bending_2013, mell_fluctuation_2015}. Adding cholesterol to the lipid composition roughly doubles  \(b=\SI{1.2 +-0.2 e9}{\newton\second\per\meter^3}\) and triples \(\kappa_c=\SI{19.2 +-0.8}{}~k_BT\) consistent with literature results \cite{rodriguez-garcia_bimodal_2009}.  As expected, the floppy GUV shows a low surface tension \(\sigma=\SI{5.1 +- 8.4e-8}{\newton\per\meter}\) (comparable to our tension detection limit), while the tense GUV of the same composition shows a roughly \(100\times\) larger value \(\sigma=\SI{4.5 +-0.1e-6}{\newton\per\meter}\). 

To better understand the complex kinematics of a nanorod moving on membranes with different tensions, we constructed a model GNR-membrane system \cite[Fig. S9]{noauthor_see_nodate-1} based on the dynamically triangulated Monte Carlo (DTMC) method \cite{ho_self-avoiding_1989}. 
The model system consists of a patch of membrane bounded by a square frame of length \(L=\SI{500}{\nano\meter}\) and a GNR represented by an ellipsoidal nanoparticle with nominal dimensions. 
This mesoscale membrane model uses a continuum approximation of bilayer membranes and computes the elastic energy using  Canham-Helfrich Hamiltonian \(H_{elastic}=\int{ds[\kappa_c(c_1+c_2)^2}/2]\) where \(c_1\) and \(c_2\) are the mean curvatures at each point on the membrane \cite{helfrich_elastic_1973}.
For the discretization of elastic energy we follow the method developed by Ramakrishnan {\it{et al.}}~\cite{ramakrishnan_monte_2010}.
In the model system, the GNR is bound to the membrane through a truncated Lennard-Jones potential (\(H_{LJ}\)) between discrete points on the membrane and nanoparticle surfaces. 
The stochastic movement of the GNR is captured by  translational and rotational Monte Carlo moves \cite{noauthor_see_nodate-1}. 

\begin{figure}
 	\centering
		 \includegraphics[width=0.49\textwidth]{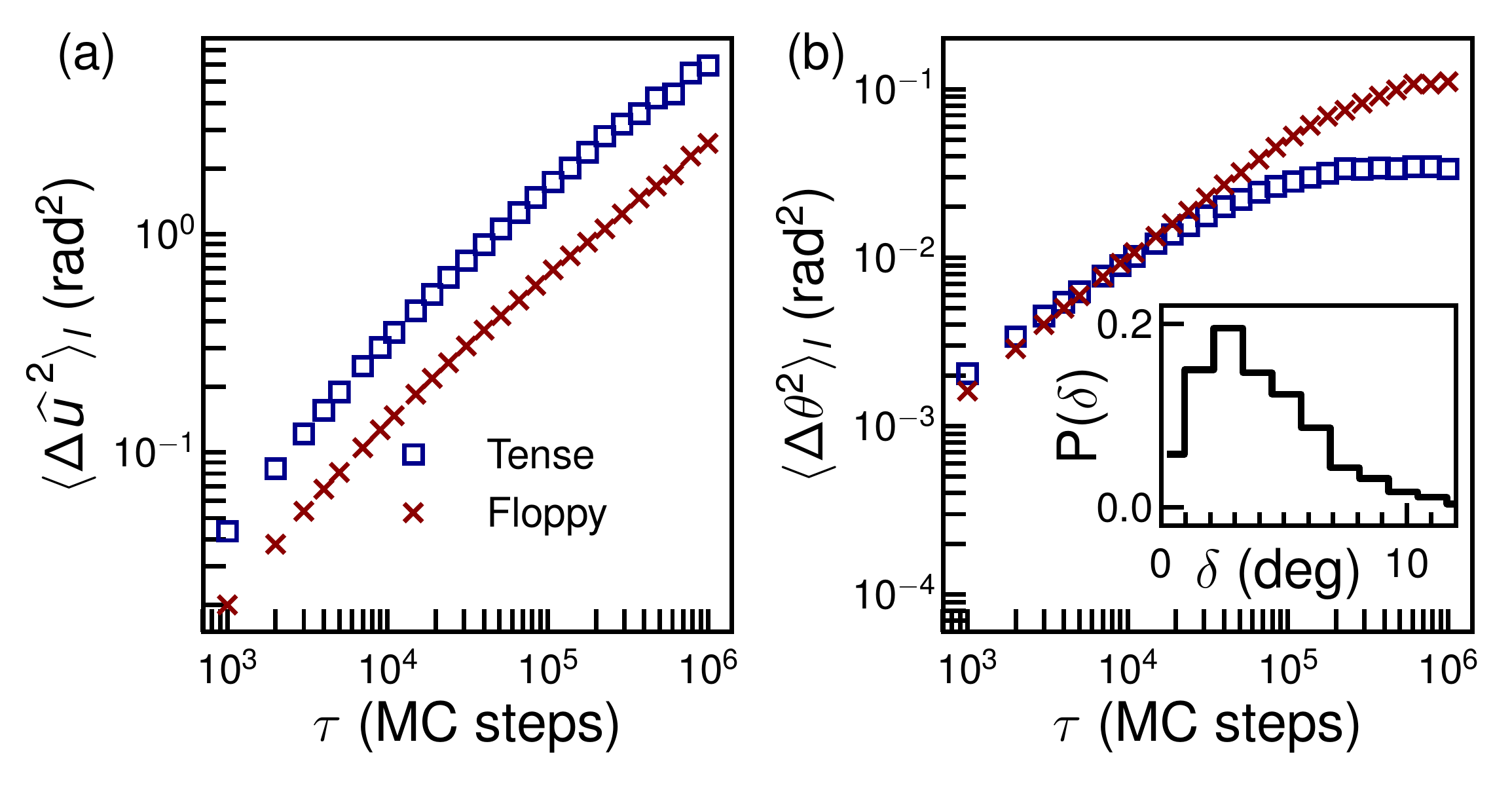}
 			\caption{Simulated motion of a nanoprobe on floppy (high excess area) and tense (low excess area) membranes. (a) MSAD of the in-plane motion of the GNR shows hindrance by membrane undulations in the floppy membrane.  (b) MSAD of the out of plane motion of the GNR shows enhancement on the floppy membrane due to higher membrane normal fluctuations.  inset: histogram of the angular difference between the simulated membrane normal \(\hat{\vec{n}}\) and that inferred from the nanoprobe orientation.}
	 \label{MC_fig}
\end{figure}

The MSAD and MSD of the out of plane motion of the simulated nanoprobe are shown in \cref{MC_fig}, for two values of excess membrane area, corresponding to `tense' and `floppy' cases \cite[Table S1]{ramakrishnan_excess_2018, noauthor_see_nodate-1}).
As seen in experiments, the simulated in-plane MSAD of the nanorod anchored on the floppy membrane is \(\approx 3\) fold smaller than that of nanorod on the tense membrane, \cref{MC_fig}(a). Examining the simulated membranes \cite[moveis S2, Fig. S7]{noauthor_see_nodate-1} reveals that the nanorod on the floppy membrane is transiently caged in the dynamically evolving valleys of the undulatory membrane leading to a suppression of its in-plane rotational motion, \cref{MC_fig}(a). Such a caging effect not only suppresses the motion, but also causes the \(\langle\Delta \hat{\vec{u}}^2\rangle\) to be slightly sub-diffusive, consistent with experiments.  As expected, the out of plane motion of the nanorods on the floppy membrane has a larger asymptote compared to the one on the tense membrane, \cref{MC_fig}(b), due to the larger total amplitude of undulations in the less tense membrane. Moreover, the time-dependent angle \(\delta\) between the simulated membrane normal and that inferred from the nanorod motion in the manner we used in experiments confirms the validity of our approach, they differ by just a few degrees.
 
Rather than GUVs made of purified lipids, the same approach can be used to study the mechanics of the plasma membrane of cultured cells.  GNRs bound to Huh7 cells perform 2D random walks on the membrane and stay in focus for more than 5 minutes, \cite[Fig. S13]{noauthor_see_nodate-1}(unlike rods in the buffer or engulfed by the cell, that rapidly go out of focus).  The orientation data resembles rapid diffusive rotation \cite[Fig. S14]{noauthor_see_nodate-1} in a plane tilted by  \(\SI{27}{\degree}\) with respect to the focal plane. Once the mean normal to the plane, \(\bar{\vec{n}}\), is found, the  time-dependent membrane normal vector, \(\hat{\vec{n}}(t)\), shown in \cref{MSD}(b), can be computed as before. This can then be decomposed into in-plane, \(\Delta\theta_\parallel\), and out of plane, \(\Delta\theta_\perp\), angular displacements, where \(\theta_\perp=cos^{-1}\hat{\vec{n}}\cdot\bar{\vec{n}}\).  

\begin{figure}
 	\centering
		 \includegraphics[width=0.49\textwidth]{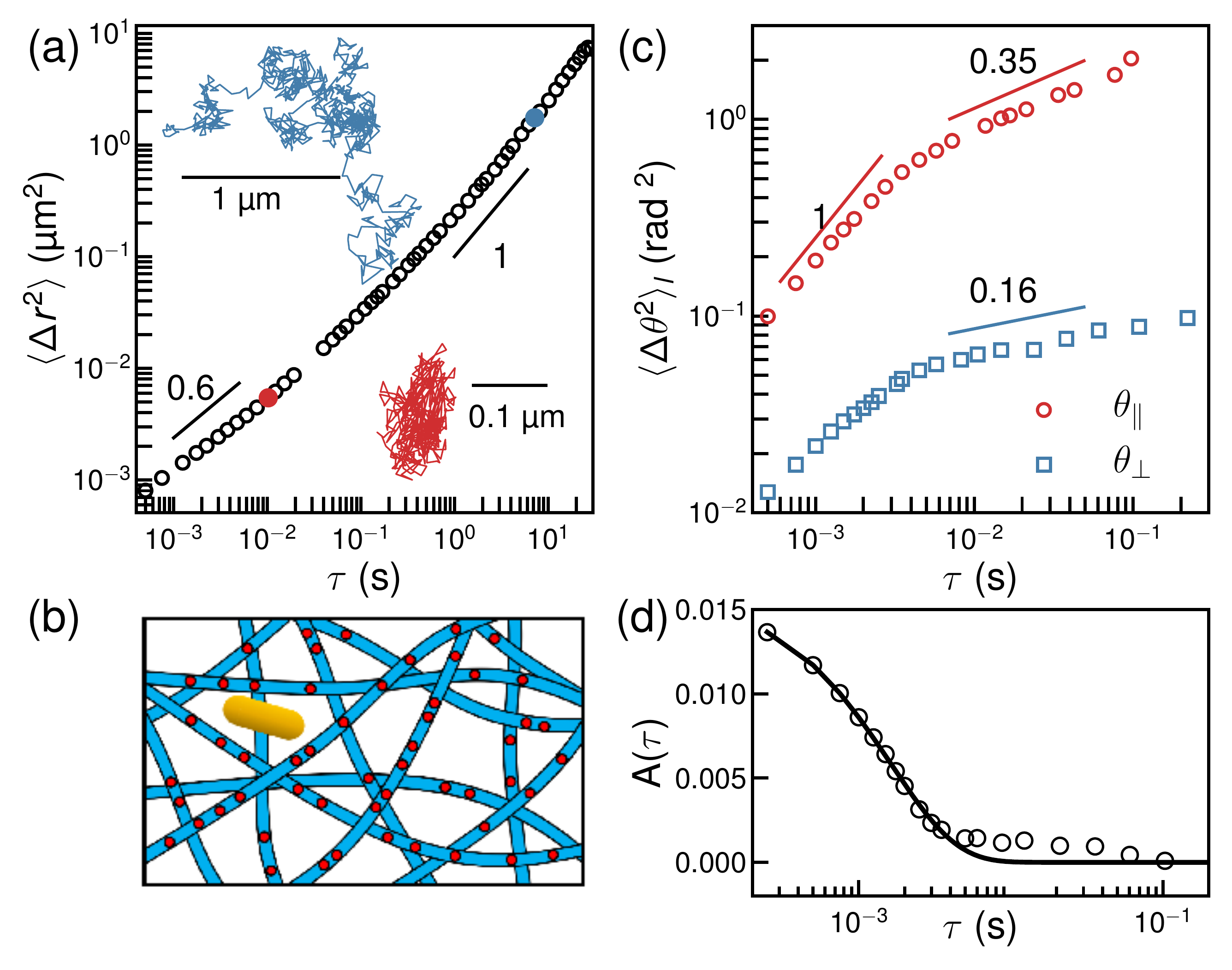}
 			\caption{ GNR tracking on a cell plasma membrane reveals translational caging and rapid normal vector fluctuations. (a) MSD of the GNR shows sub-diffusive caging at short lag-times,  and diffusion at long lag-times, illustrated by the red and blue trajectories spanning \SI{10}{\milli\second} and \SI{7}{\second} respectively. Eye guides show the logarithmic slope. (b) Schematic of a GNR caged by integrin proteins (red circles), anchored to actin filaments under the membrane. (c) MSAD of the in-plane and out of plane motion of the GNR show rapid motion at short times, and caging at long lag times resembling cell cortical fluctuations. (d) The covariance of the out of plane angle reveals the rapid decay of plasma membrane normal fluctuations relative to the cortex.} 
	 \label{Cell_Exp}
 \end{figure}

The MSD and a typical trajectory of the GNR at short lag times, \cref{Cell_Exp}(a), reveals subdiffusive motion of GNR.  The plasma membrane is tethered to underlying actin filaments by integrin proteins which form a structure akin to `picket fences' surrounding `corrals', \cref{Cell_Exp}(b), \cite{hamley_biological_2007, kusumi_paradigm_2005, morone_three-dimensional_2006}. We can interpret the MSD as due to cage diffusion and escape from cages/corrals of variable sizes. The MSD crosses over to purely diffusive motion with \(D_t=\SI{0.06+-0.01}{\micro\meter^2\per\second}\) at a lengthscale of \((\SI{200}{\nano\meter})^2\), comparable to the largest expected corral size \cite{mcmahon_membrane_2015}. 

Both \(\langle\Delta\theta^2_\parallel\rangle\) and \(\langle\Delta\theta^2_\perp\rangle\) show complex lag-time dependence, \cref{Cell_Exp}(c). As in the GUV, the out of plane fluctuation is much slower than the in-plane diffusion of the GNR. At short lag-time, \(\tau<\SI{10}{\milli\second}\), the \(\theta_\parallel\) shows a pure diffusive behavior, with the diffusivity \(D_{r}=\SI{48.0 +- 6.4}{rad^2\per\second}\) comparable with that in a tense pure lipid membrane. At intermediate lag time \(\tau>\SI{10}{\milli\second}\), however, the rotational motion is subdiffusive consistent with angular caging.

Due to its coupling to the underlying actomyosin cortex, the undulation dynamics of the plasma membrane on long length and timescales is that of the cortex \cite{doherty_mediation_2008}.
Indeed, the sub-diffusive exponent of \(\langle\Delta\theta^2_\perp\rangle\) at \(\tau>\SI{10}{\milli\second}\) is consistent with expectations for cortical fluctuations, since the cortex has been shown to have a dynamic shear modulus varying as \(\sim \omega^{0.16}\)\cite{hoffman_cell_2009}. At \(\tau<\SI{5}{\milli\second}\), \(\langle\Delta\theta^2_\perp\rangle\) increases subdiffusively. We can isolate this short-time motion of the membrane by computing the covariance of the out of plane angle of GNR, \(A(\tau)=\langle \theta_\perp(t+\tau)\theta_\perp(t) \rangle_t\), \cref{Cell_Exp}(d), which decays exponentially with a relaxation rate of \(\omega = \SI{610 +- 24}{\per\second}\). 

Hypothesizing that the short time dynamics of the normal vector is due to plasma membrane undulations, we can create a simple model by integrating Eqn. \ref{MSAD_normal_eq} over a narrow range of wave-numbers, \(\delta q\), set by the corral size and the nanorod length, yielding a single exponential decay \(A(\tau)=(k_BT / 2\pi\kappa_c)(\delta q / q)e^{-\omega_1\tau}\) consistent with our observations, \cite{noauthor_see_nodate-1}. Using an estimated value of  \(\delta q/q=0.35\), this provides an estimate of the bending rigidity of the plasma membrane, \(\kappa_c\approx \SI{3.5}{}~k_BT\), roughly a factor of 5 smaller than expected \cite{steinkuhler_mechanical_2019}.  

This too small inferred bending modulus can be readily explained by additional sources of undulation not accounted for by the simple model.  For one, the integrins presumably perturb the membrane height, producing a corrugation that leads to dynamic out of plane motion as the nanorod diffuses over it.  For another, the presence and diffusion of membrane curvature inducing proteins would also increase the undulation amplitude \cite{tourdot_application_2015}.  Sorting out these multiple contributions will require improved models and future experiments that vary \(\delta q / q\) by using different length nanorods.  
 
 In conclusion, the nanorod tracking approach we present here enables the reliable measurement of membrane properties of cell-sized GUVs and the undulation dynamics of the plasma membranes of single living cells. Notably, the measurements here are not near their physical limits (e.g. due to photodamage or heating), and so should be readily extendable using faster cameras and brighter laser illumination.  This will allow the use of still smaller GNRs, and thus the optical measurement of undulations having wavelengths far smaller than the diffraction limit. Since the orientational tracking is polarimetric and can be performed at low magnification, a wide-field camera would also enable high-throughput measurements of different cells or cell regions simultaneously. Last, our approach promises to enable the study of the mechanics of the membranes of procaryotes, sub-cellular organelles or the dynamics of cell-cell junctions. 

\vspace{3mm}
\begin{acknowledgments}
This work was supported by the Penn PSOC program, NIH U54-CA193417, and partial support from Grant No. 55120-NDS from the ACS Petroleum Research Fund. 
\end{acknowledgments}

\bibliography{Ref_GUV.bib}
\end{document}



\title{Supplementary Material for \\ Probing lipid membrane bending mechanics using gold nanorod tracking}


\author{Mehdi Molaei} 
\affiliation{Chemical and Biomolecular Engineering, University of Pennsylvania, Philadelphia, PA 19104}

\author{Sreeja Kutti Kandy} 
\affiliation{Bioengineering, University of Pennsylvania, Philadelphia, PA 19104}

\author{Zachary T. Graber} 
\affiliation{Chemistry, University of Pennsylvania, Philadelphia, PA 19104}

\author{Tobias Baumgart}
\affiliation{Chemistry, University of Pennsylvania, Philadelphia, PA 19104}

\author{Ravi Radhakrishnan}
\affiliation{Bioengineering, University of Pennsylvania, Philadelphia, PA 19104}

\author{John C. Crocker}
\affiliation{Chemical and Biomolecular Engineering, University of Pennsylvania, Philadelphia, PA 19104}

\maketitle

\section{Experimental Methods}
\subsection{Sample Preparation}

{\it{GNR solution:}}
The carboxyl functionalized GNR solution was purchased from Nanopartz (A12-40-850-TC) with GNR concentration of \SI{1.1e12}{nps\per\milli\liter}.
In a \SI{1.5}{\milli\liter} microcentrifuge vial, we mixed \SI{20}{\micro\liter} of this GNR solution with \SI{0.98}{\milli\liter} PBS buffer with pH of 7.4 to make a dilute GNR suspension. 
Magainin 2 peptide was purchased from Genscripts (RP11232-0.5). \SI{1}{\milli\liter} DI water was added to \SI{10}{\milli\gram} magainin 2 to make a \SI{4}{\milli\mole} peptide solution.
To functionalize GNRs with the peptide molecules, we used a protocol designed to conjugate nanoparticles with nuclear localization signal (NLS) peptide \cite{xie_nuclear_2009} with minor alterations.  
\SI{1}{\m l} DI water was added to \SI{.19}{\milli\gram} of N-ethyl-N'-dimethylaminopropyl-carbodiimide (EDC, from Sigma-Aldrich E6383-1G) to make \SI{1}{\milli\mole} EDC solution.
In a separate vial, \SI{1}{\milli\liter} DI water was added to \SI{.11}{\milli\gram} of N-Hydroxysuccinimide (NHS, from Sigma-Aldrich 130672-5G) to yield \SI{1}{\milli\mole} NHS solution.   
After preparing these samples, we added \SI{10}{\micro\liter} of the EDC solution and \SI{5}{\micro\liter} of the NHS solution to the dilute GNR suspension and vortexed for 10 seconds. Then, we added \SI{20}{\micro\liter} of the peptide solution to the sample. 
The sample was stirred for 48 hours. At the end, the sample was centrifuged at 1000 rcf and resuspended in the PBS buffer twice to remove the excess peptides. To verify that the GNRs were not aggregated throughout the process, the uv-vis absorbance of the final sample was measured and compared with the absorbance of the original sample.

{\it{GUV solution:}}
Giant Unilamellar Vesicles (GUVs) were prepared from lipid stocks of 1,2-dioleoyl-sn-glycero-3-phosphocholine (DOPC), 1,2-dioleoyl-sn-glycero-3-phospho-L-serine (DOPS), and cholesterol using the standard method of electroformation \cite{angelova_liposome_1986, tian_sorting_2009}. Briefly, the lipid stocks were prepared in organic solvent with the desired lipid composition and spread on indium tin oxide (ITO) coated slides. To remove all traces of organic solvent and form a dry lipid film, the lipid-coated slides were placed under vacuum for two hours. An electroformation chamber was formed using two lipid-coated ITO slides with \SI{0.8}{\milli\meter} rubber spacers. Inside the electroformation chamber the lipid film was hydrated with 400-450 \SI{}{\micro\liter} of \SI{0.3}{M}  sucrose. Electroformation was performed by applying a \SI{10}{\hertz}, 2-4 Vpp electric field over two hours to form the final GUV dispersion.
To perform the experiment, we add \SI{10}{\micro\liter} of the peptide coated GNR solution to \SI{200}{\micro\liter} of the GUV solution. The mixed solution was loaded between two \#1.5 coverslips and is sealed carefully with vacuum grease.

{\it{Huh7 cells:}}
Huh7 cells were cultured following the standard protocol. Cells were grown in a \SI{37}{\degree C} incubator under 5\% CO2. They are cultured in 35 mm dish with \#1.5 coverslip in the middle (Mattek, P35GCOL-1.5-10-C) in \SI{2}{\milli\liter} of DMEM with  10\% FBS and penicillin/streptomycin. Cells were grown to \(\sim\)50\% confluence which took about 2 days. Before imaging, the culture medium were replaced with a PBS buffer with pH 7.3, and \SI{10}{\micro\liter} GNR solution was added.  

\subsection{Imaging and particle tracking}
Highly polarized light scattered from GNRs are imaged using a custom-built laser-illuminated dark field microscope. The detail of the microscope is provided in an earlier publication \cite{molaei_nanoscale_2018}. We use an 100X oil-immersion objective (Leica HCX PL APO 100X-1.4) and high speed CMOS Phantom IV camera (Vision Research) to collect and simultaneously image the light in two orthogonally polarized channels with effective pixel size of \(\SI{0.1}{\micro\meter}\). All the imaging has been performed at room temperature.

\begin{figure}
  \centering
    \includegraphics[width=0.4
    \textwidth]{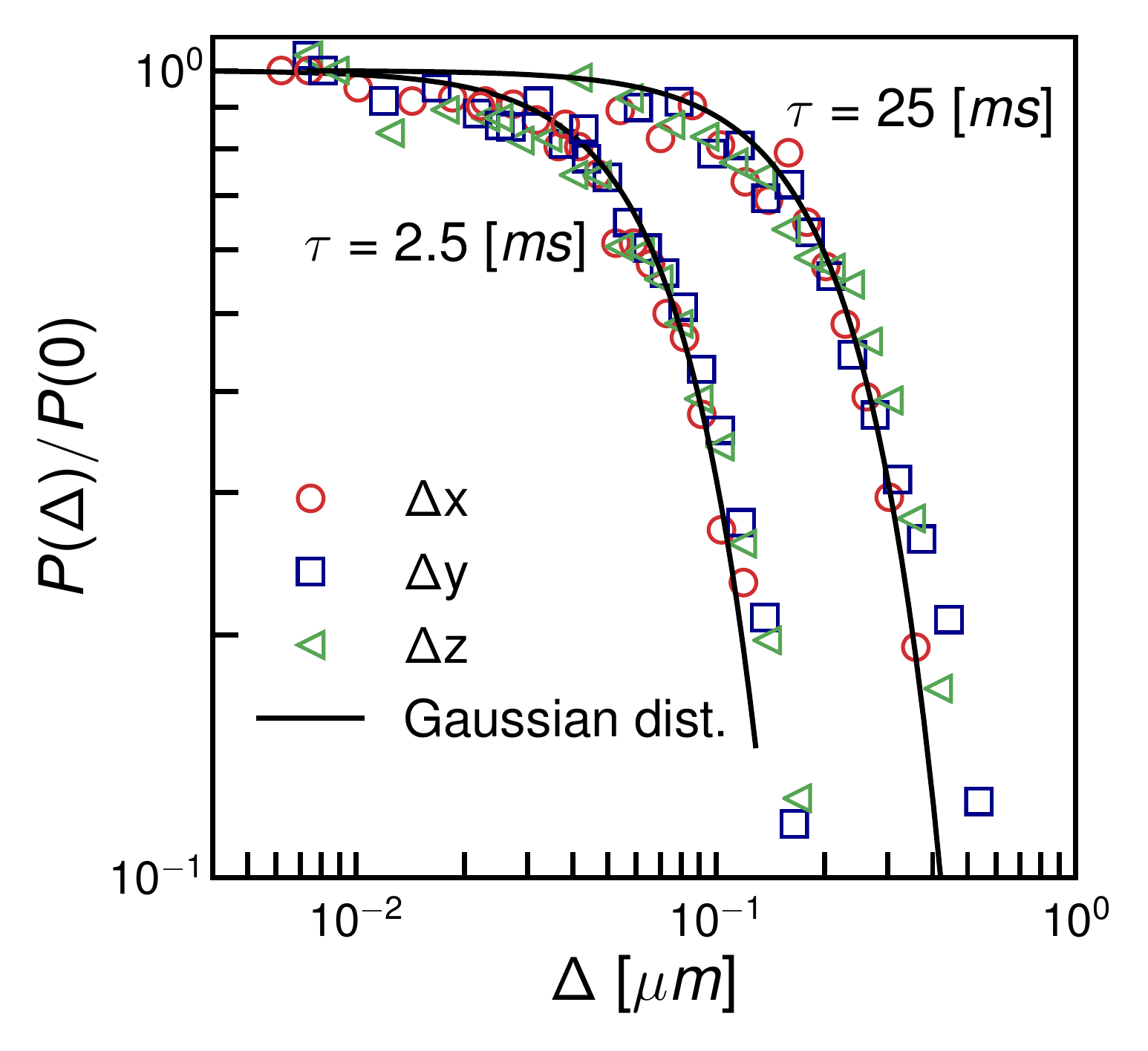}
    \caption{The distribution of displacements in all three direction, or van Hove functions for \(\tau=\SI{2.5}{\milli\second}\) and \(\tau=\SI{25}{\milli\second}\), follow normal distributions.}
	\label{VanHove}
 \end{figure}

The lateral motion, xy, of the GNR is tracked using the centroid of the DFM images. The center \(x_c,y_c\) and  radius, \(R\), of the GUV are determined by the phase contrast microscopy. The position of GNR in z direction is then computed through \begin{equation}
{z=[R^2-(x-x_c)^2-(y-y_c)^2]^{\frac{1}{2}}}.
\end{equation}
The probability of the GNR displacement in lateral and in-depth direction \cref{VanHove} shows isotropic diffusion process with Gaussian distributions insuring the accuracy of 3-d trajectories. The MSD of a particle diffusing over the surface of a sphere follows
\begin{equation}
    \label{MSD_eq}
    \langle\Delta \vec{r}^2 (\tau)\rangle=2R^2(1-exp[-\frac{2D_t \tau}{R^2}]),
\end{equation}
where \(D_t=k_BT/\gamma_t\) is the translational diffusion coefficient, and \(\gamma_t\), \(k_B\), and \(T\) are the translational drag coefficient, Boltzmann constant, and the temperature. For \(\tau\ll D_t/R^2\), \cref{MSD_eq} simply becomes a Brownian random walk on a 2D plane with \(\langle\Delta \vec{r}^2 (\tau)\rangle=4D_t\tau\).  

\section{Rotational diffusion of a rod on a tilted plane}

\begin{figure}
  \centering
    \includegraphics[width=.85
    \textwidth]{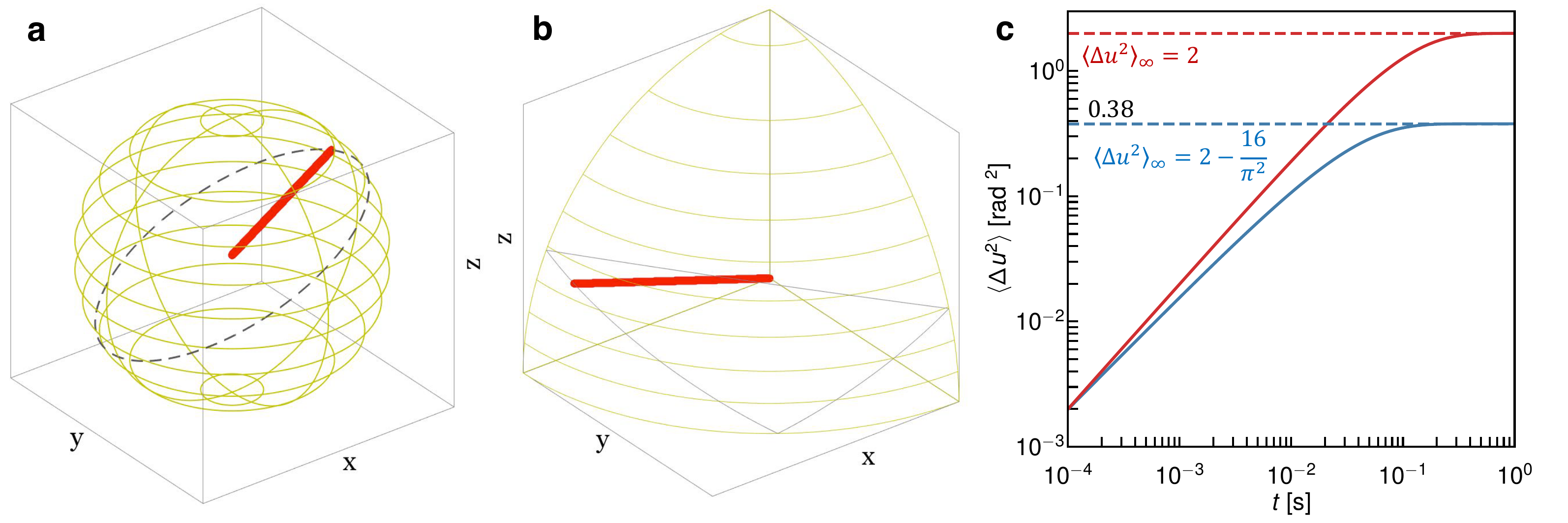}
    \caption{Motion of rod on a tilted plane with \(\beta=\SI{30}{\degree}\) and \(\phi=\SI{45}{\degree}\). a) \(\hat{\vec{u}}\), the unit vector pointing the major axis of the rod, red thick line, follows a circular path shown with gray dashed lines. b) Mapped orientation vector, \(|\hat{\vec{u}}|\) follows a peculiar loop in octant of sphere shown by grey line.  c) MSAD of the unite vector \(\hat{\vec{u}}\) red line, and its mapped on the octant, \(|\hat{\vec{u}}|\) blue line follow exponential and stretched exponential functions respectively.}
	\label{Circle}
 \end{figure}

Consider a rod  anchored on a 2D plane with normal vector of 
\begin{dmath}
\hat{n}=sin\beta cos\phi \hat{e}_x+sin\beta sin\phi \hat{e}_y+cos\beta\hat{e}_z,
\end{dmath} 
where \(\phi\) and \(\beta\) are the azimuthal and polar angles. Now let the rod rotationally diffuse in this 2D plane with diffusion coefficient of \(D_r\). The unit vector of the rod mapped on the plane is \(u'=cos\theta\hat{e}'_x+sin\theta\hat{e}'_y\) where \(\hat{e}'_{x,y}\) are the unit vectors of the coordinate system aligned on the 2D plane and \(\theta\) is generated by random rotational Brownian motion where \(\langle \Delta\theta^2(t)\rangle=2D_rt\). It can be shown that the unit vector of the rod  is
\begin{dmath}
\hat{\vec{u}}=[cos\theta-(1-cos\beta)cos\phi cos(\theta-\phi)]\hat{e}_x+[sin\theta-(1-cos\beta)sin\phi sin(\theta-\phi)]\hat{e}_y-sin\beta cos(\theta-\phi)]\hat{e}_z,
\end{dmath} 
which follows a circular loop. For instance, if the rod diffuses on a plane with \(\beta=\SI{30}{\degree}\) and \(\phi=\SI{45}{\degree}\), the unit vector of the rod will follow the circular path shown in \cref{Circle}(a).

If we were measuring the orientation of this rod with our imaging technique, because \(\hat{\vec{u}}\) is mapped to an octant of a sphere, the measured orientation vector would have followed a path shown in \cref{Circle}(b).
Video 1 respectively show the random motion of the the rod diffusing on this plane and its mapped orientation to the octant.
The MSAD of the fully resolved unit vector is an exponential function, \(\Delta \hat{\vec{u}}^2(\tau)=2(1-exp[-D_rt]) \), and MSAD of the measured unit vector follows an stretched exponential function 
\begin{equation}\label{MSAD_str}
    \langle\Delta \hat{\vec{u}}(\tau)^2\rangle=\Delta \hat{\vec{u}}_\infty^2(1-exp[-(kD_r\tau)^\xi]),
\end{equation}
as shown in \cref{Circle}(c), where \(k=4.04\) and \(\xi=0.93\) are the parameters of the stretched exponential, and \(\Delta \hat{\vec{u}}_\infty^2\) is the asymptote value.
For a point moving randomly on a circle, the asymptote value of the MSD is calculated as
\begin{equation}
\Delta \hat{\vec{u}}_\infty^2=\frac{1}{4\pi^2}\int_0^{2\pi}{\int_0^{2\pi}{[1-cos(x_1-x_2]dx_1dx_2}}=2.
\end{equation}
For a point moving moving on an arc of size \(\pi/2\), the asymptote  is calculated as
\begin{equation}
\Delta \hat{\vec{u}}_\infty^2=\frac{1}{4\pi^2}\int_0^{\pi/2}{\int_0^{\pi/2}{[1-cos(x_1-x_2]dx_1dx_2}}=2-\frac{16}{\pi^2}.
\end{equation}
Simulation of the rod on planes with any polar and azimuthal angles shows that the MSAD of the unit vector does not depend on the orientation of the plane. 

\begin{figure}
\centering
    \includegraphics[width=.4\textwidth]{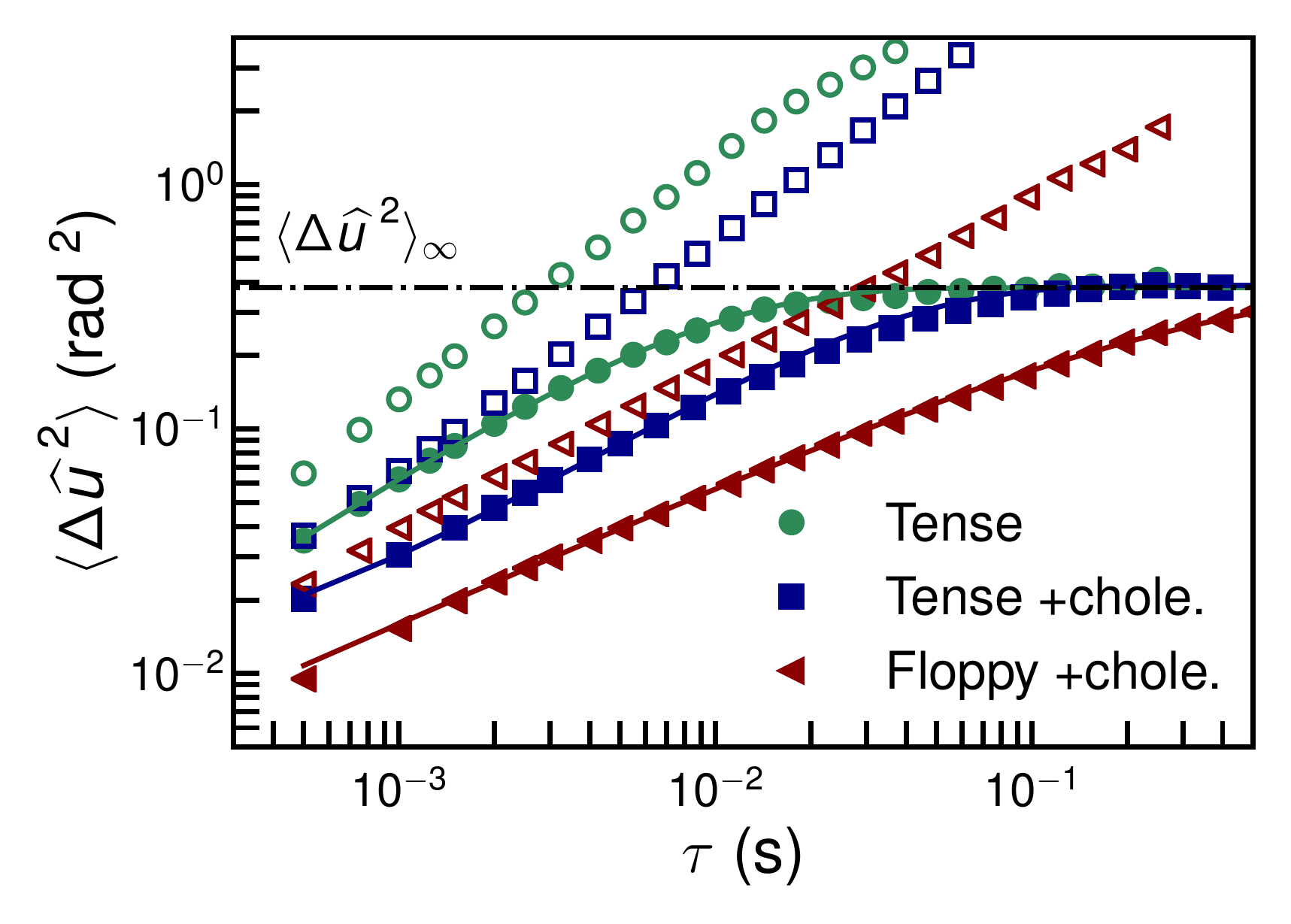}
    \caption{ Measured bounded MSADs, closed symbols, and scaled unbounded MSADs, open symbols, of the GNRs on GUVs. Lines are fits to \cref{MSAD_str}.}
	\label{MSAD_single}
\end{figure}

We can compute a lag-time independent mapping between this bounded MSAD and an unbounded MSAD corresponding to an orientation vector which is not limited to the unit octant, 
\begin{equation}
    \langle\Delta \hat{\vec{u}}^2(\tau)\rangle_l=\frac{4}{k}\left[-\ln\left( 1-\frac{\langle\Delta \hat{\vec{u}}^2(\tau)\rangle}{\Delta \hat{\vec{u}}_\infty^2} \right) \right]^{\frac{1}{\xi}}.
\end{equation}
Measured bounded MSADs of the GNRs on tense and floppy GUVs and their corresponding scaled unbounded MSADs are shown in \cref{MSAD_single}. The unbounded MSADs are reported in the main text, Fig. 2(e). 

\begin{figure}
\centering
    \includegraphics[width=.4\textwidth]{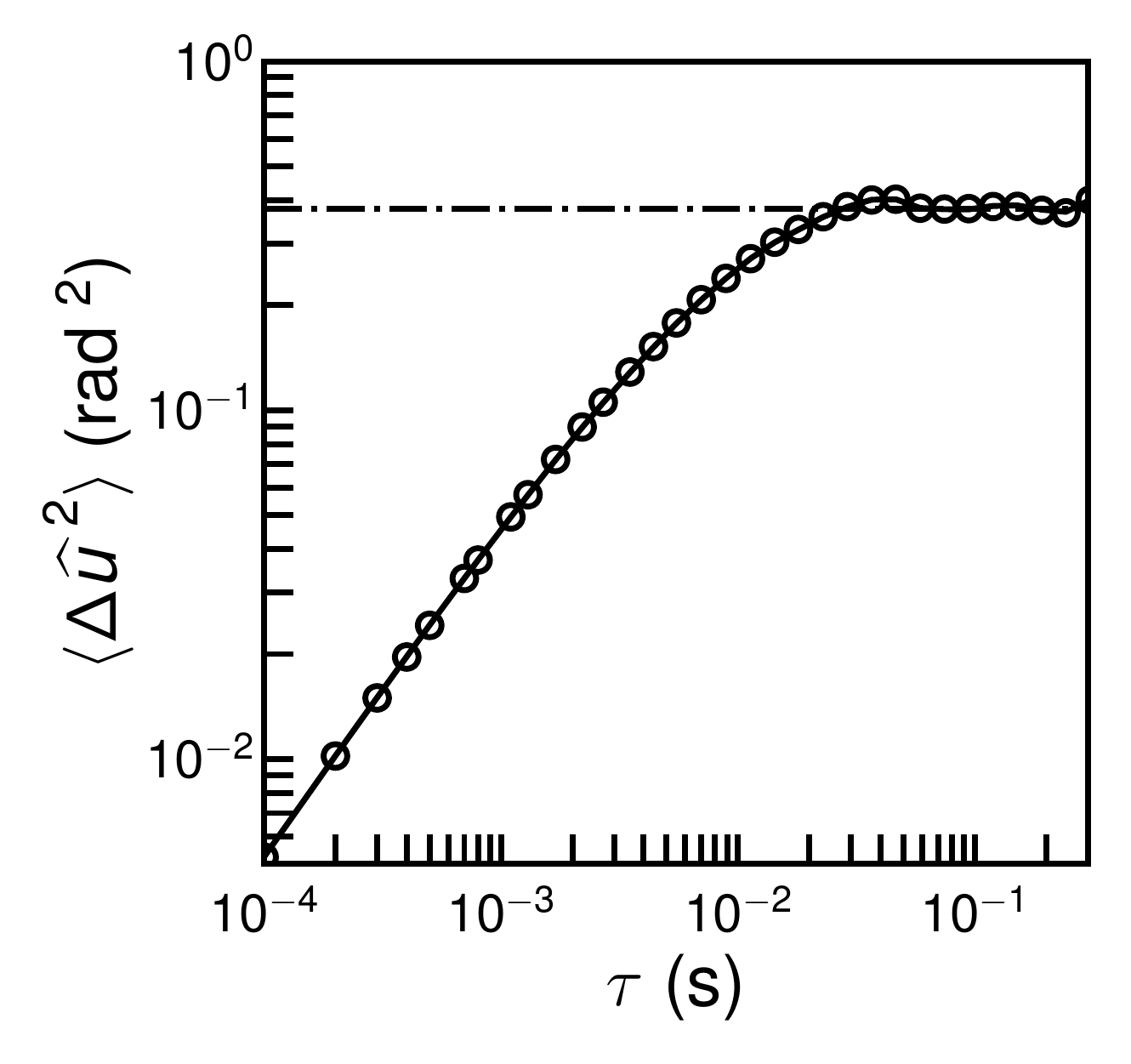}
    \caption{MSAD of random motion of a nanorod with in-plane rotational diffusion of \(D_r=\SI{27}{rad^2\per\second}\) and translational diffusion of \(D_t=\SI{1}{\micro\meter^2\per\second}\), symbols, and \(D_t=0\), solid line.  The GUV radius is \(R=\SI{7}{\micro\meter}\), similar to the experiment.}
	\label{MSADonGUV}
\end{figure}

In the main text, we have associated \(\langle\Delta \hat{\vec{u}}^2(\tau)\rangle\) solely with the in-plane rotational motion of the GNR, but one could argue that the translational motion of a GNR on the GUV can also contribute to the change in the orientation vector. However, since \(D_t/R^2\ll D_r\) we can confidently neglect such contribution. Nevertheless, we have generated a random rotational motion of a GNR on a GUV with and without random walk. The result shown in \cref{MSADonGUV} confirms our assessment. In a similar way, it can be shown that out of plane undulation of the membrane is not large enough to contribute to \(\langle\Delta \hat{\vec{u}}^2(\tau)\rangle\). 

\section{Mobility of the rod associated with membranes}
\begin{figure}
    \centering
    \includegraphics[width=0.85\textwidth]{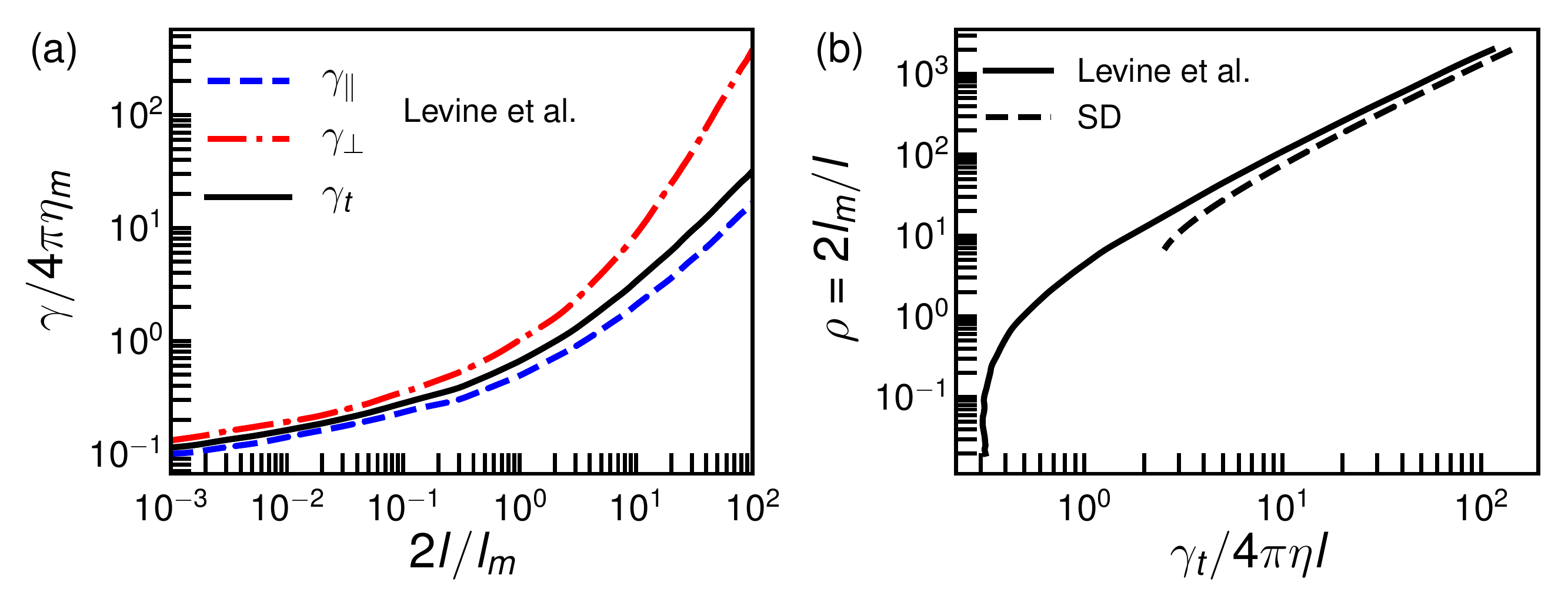}
    \caption{a) Parallel, \(\gamma_\parallel\), and perpendicular, \(\gamma_\perp\), drag coefficient of a rod with aspect ratio of 3 diffusing on a viscous membrane obtained from Levine \textit{et al}. \cite{levine_dynamics_2004} and the total translational drag coefficient, \(\gamma_t=2\gamma_\parallel\gamma_\perp/(\gamma_\parallel+\gamma_\perp)\). b) Relation between dimensionless number \(\rho=2\eta_m/l\eta=2l_m/l\) and the translational drag coefficient of the rod, evaluated from \(\gamma_t\) in panel (a) and of the disk like object with diameter l predicted by Saffmann and Delbr\"uk (SD) model \cite{saffman_brownian_1975}.}
    \label{dragcoef}
\end{figure}

The drag coefficient of the GNR depends on the 2D membrane viscosity, \(\eta_m\), bulk fluid viscosity \(\eta\), and the geometry of the rod. We first start by using the theoretical approach developed by Levine \textit{et al}. \cite{levine_dynamics_2004} to calculate the translational drag coefficients parallel, \(\gamma_\parallel\), and normal, \(\gamma_\perp\), to the long axis of a rod on the viscous membranes. In 2D membrane total translational drag coefficient can be estimated as \(\gamma_t=\frac{2\gamma_\parallel\gamma_\perp}{\gamma_\parallel+\gamma_\perp}\), \cref{dragcoef}(a). 
From this function, we establish an explicit relation between the membrane viscosity and translational drag coefficient in reduced units, \cref{dragcoef}(b). Using the nominal length and the aspect ratio of the GNR, we obtain a rough estimate of membrane viscosity from the translational drag coefficient. For no cholesterol GUV, we had measured \(D_t=\SI{0.52}{\micro\meter^2\per\second}\) which yields \(\gamma_t=k_BT/D_t=\SI{7.8}{\nano\pascal\second\meter}\). Using numerically evaluated function shown in \cref{dragcoef}(b), we estimate \(\rho=2l_m/l=38\). In the limit that \(a<l \ll l_m\), the drag coefficient of the rod become independent of the rod orientation and aspect ratio \cite{levine_dynamics_2004}. 
Hence, we simply use Saffmann and Delbr\"uk (SD) model for rotational and translation drag coefficients \cite{saffman_brownian_1975}  with the correction factor proposed by Naji \textit{et al}. \cite{naji_corrections_2007}
\begin{equation}
    \gamma_t=2\pi\eta l[\frac{\rho}{ln \rho-\gamma_E}+c/4\pi],
\end{equation}
\begin{equation}\label{eq:gr}
    \gamma_r=\frac{\pi\eta l^3}{2} \rho,
\end{equation}
where, \(\gamma_E\) is Euler-Mascheroni constant, \(\rho=2\eta_m/l\eta\) is the dimensionless length scale indicating the importance of membrane mechanics to bulk hydrodynamic, and parameter \(c\approx \pi (a/l)^2\) is roughly the volume of bulk fluid displaced by the membrane deformation to \(l^3\) and is expected to be of order 1 \cite{hormel_measuring_2014}. For an object diffusing on a viscous membrane, the radius of probe often is not considered as the effective radius for diffusion process; in this setting the number of bounding sites to the bilayer and also distortion induced by the object determined the effective radius \cite{hormel_measuring_2014}. Because of these factors and also uncertainty in the length of GNRs, we consider the effective length of the GNR as an unknown parameter in the diffusion process. Knowing both rotational and translational drag coefficients of the GNR, however, we determine both membrane viscosities and effective size of the GNRs. Applying SD model we determine the relation between the ratio of translational to rotational drag coefficient,\(\alpha=\gamma_t/\gamma_r^{1/3}(4\pi\eta)^{2/3}\) and length scale associated with the membrane viscosity and the bulk fluid viscosity.
\begin{equation}
    \alpha=\frac{\rho^{2/3}}{ln\rho-\gamma_E}
\end{equation}
For example, for the GUV with no cholesterol,  \(\alpha=2.73\) yielding \(\rho=11.4\); placing this value in \cref{eq:gr} the effective length of the GNR is determined, \(l=\SI{202}{\nano\meter}\). The membrane viscosity is simply calculated, \(\eta_m=\rho\eta_ml/2=\SI{1.15}{\nano\pascal}\).

\section{Estimating membrane normal from orientation of GNR}
\begin{figure}
 	\centering
		 \includegraphics[width=.85\textwidth]{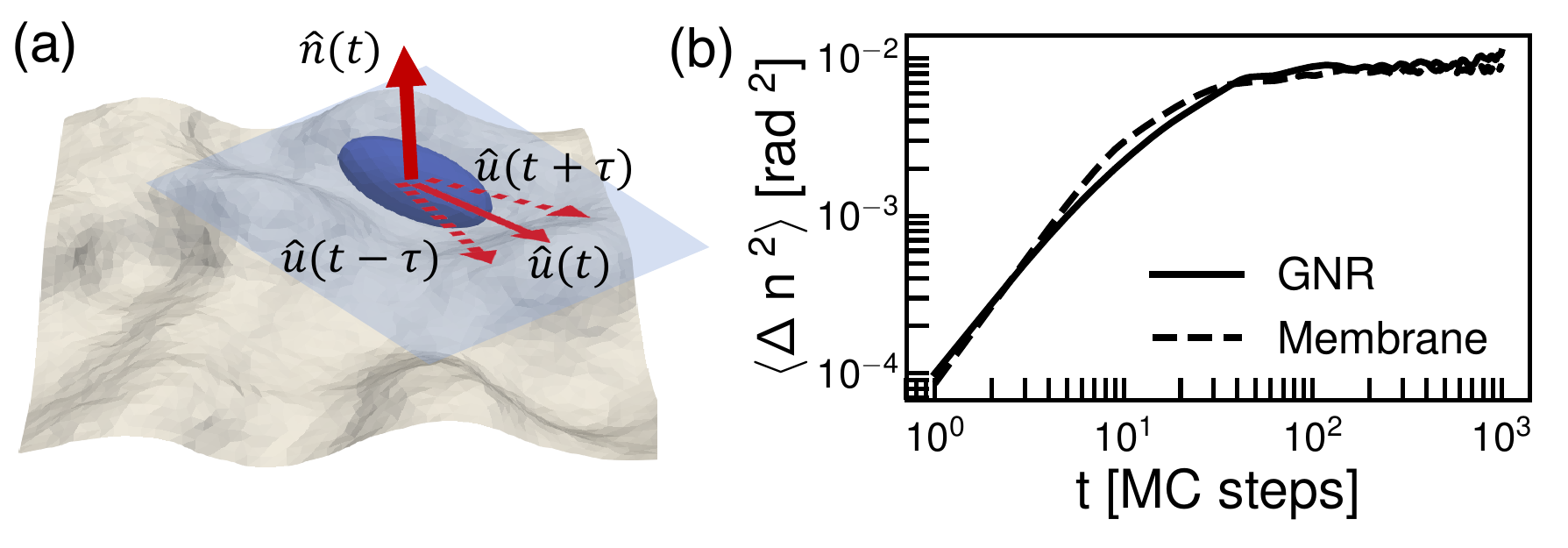}
 			\caption{a) Measurement of membrane normal vector from orientation of membrane bond GNR. b) MSAD of the membrane normal vector, \(\hat{n}_s\) dashed line, and estimated normal vector, \(\hat{n}\) solid line}
	 \label{PRL4}
 \end{figure}

We use numerically simulated motion of the nanoprobe on the model membrane. First, we compute the local normal vector of the segment of the membrane associated with the rod over time, \(\hat{\vec{n}}_s(t)\).
We then estimate the normal vector with the proposed method. The angle difference between normal vector and estimated normal \(\delta=cos^{-1}(\hat{\vec{n}}_s\cdot\hat{\vec{n}})\) indicates the agreement of the estimated values with normal vector, shown in the inset to Fig. 3(b).
Therefore, MSAD of the membrane fluctuation can be estimated by measuring MSAD of the normal vector of GNR, \(\hat{\vec{n}}\), as plotted in \cref{PRL4}(b). 

\begin{figure}
    \centering
    \includegraphics[width=.4\textwidth]{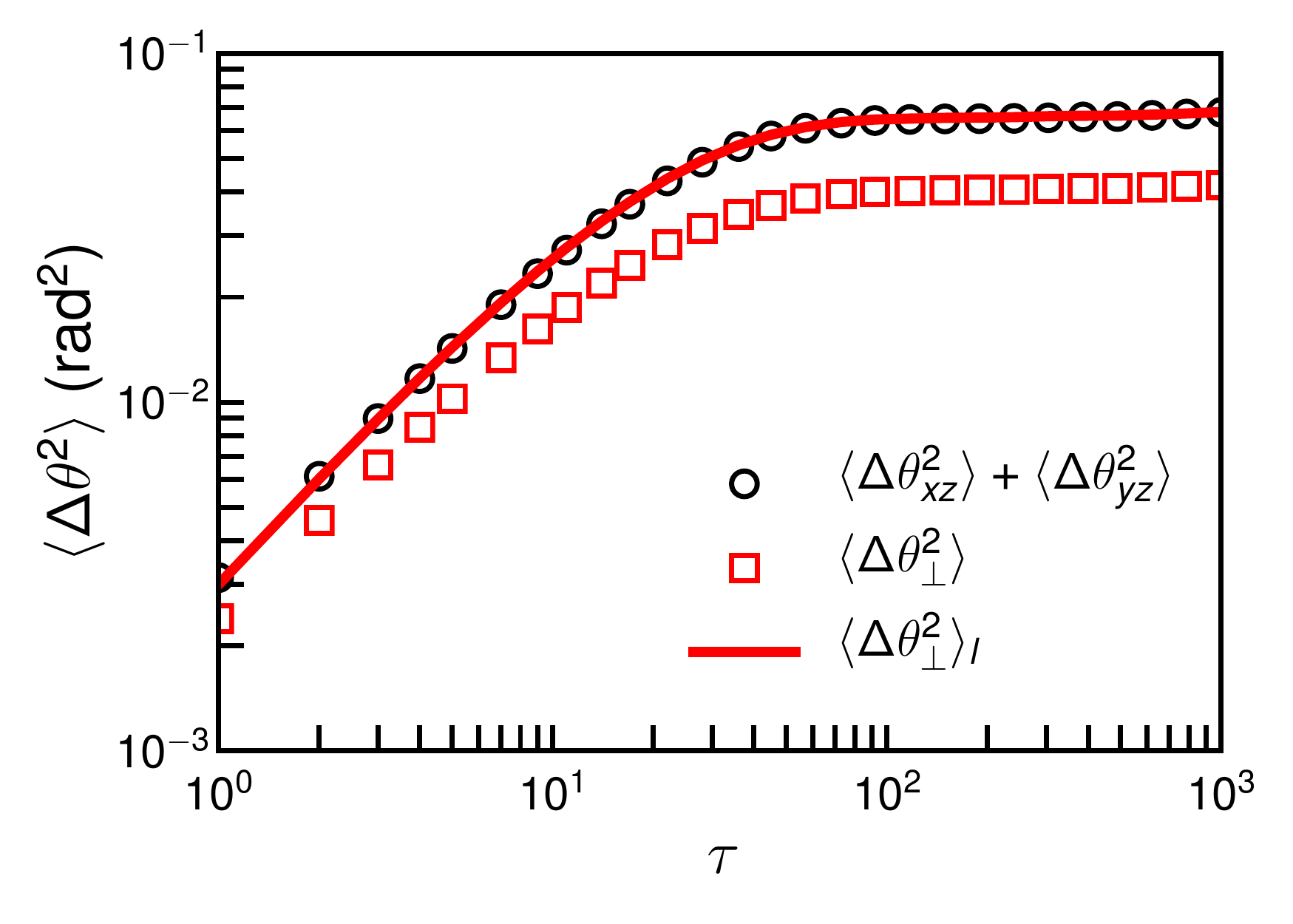}
    \caption{MSAD of the out of plane motion of a membrane with a normal vector with a random walk following Kelvin-Voigt model. The logic behind selecting this model is that the measured out of plane MSADs, Fig. 2, follow a similar MSD of colloids in a Kelvin-Voigt materials. }
    \label{MSAD_normalA}
\end{figure}

The out of plane angular motion of the GNR is then determined by tracking the angle between \(\hat{\vec{n}}\) and the average normal vector of the plasma membrane  \(\theta_\perp=\cos^{-1}\bar{\vec{n}}\cdot\hat{\vec{n}}\) or the angle between \(\hat{\vec{n}}\) and the position vector of the GNRs on the GUVs \(\theta_\perp=\cos^{-1}\hat{\vec{r}}\cdot\hat{\vec{n}}\). This process bounds all the angle information to \(0<\theta_\perp<\pi/2\). Hence, we use previously developed approach that computes a lag-time independent mapping between bounded and unbounded MSDs \cite{molaei_nanoscale_2018}. To ensure that this mapping approach faithfully map bounded \(\langle\Delta \theta^2\rangle\) to unbounded one \(\langle\Delta \theta^2\rangle_l\), we simulate a random angular walk of the normal vector. Without loss of generality, we set \(\bar{\vec{n}}=\hat{\vec{e}}_z\) and generate two sets of random walks for the angular motion of \(\theta_{zx}=\tan^{-1}(n_x/n_z)\) and \(\theta_{zy}=\tan^{-1}(n_y/n_z)\). Both of these walks are generated in a way that their MSDs follow Kelvin-Voigt model \cite{khan_random_2014} which are similar to the measured \(\langle\Delta\theta_\perp^2\rangle\). \Cref{MSAD_normalA} shows the MSAD of this randomly generated walk, \(\langle\Delta\theta_{xz}^2(\tau)\rangle+\langle\Delta\theta_{yz}^2(\tau)\rangle\). The normal vector of the membrane is constructed as
\begin{equation}
    \hat{\vec{n}}=\frac{1}{(\tan^2\theta_{xz}+\tan^2\theta_{xy}+1)^{\frac{1}{2}}}[\tan{\theta_{xz}}\hat{\vec{e}}_x+\tan{\theta_{yz}}\hat{\vec{e}}_y+\hat{\vec{e}}_z],
\end{equation}
and \(\theta_\perp=\cos^{-1}{\left(\left[\tan^2\theta_{xz}+\tan^2\theta_{xy}+1\right]^{-1/2}\right)}\). The MSAD of this out of plane angle, \(\langle \Delta \theta_\perp^2\rangle\) is shown in \cref{MSAD_normalA}. The mapped MSAD is then obtained from

\begin{equation}
    \langle \Delta \theta_\perp^2(\tau)\rangle_l=\frac{\pi}{0.85}\left[-\langle\Delta\theta^2(\infty)\rangle\ln{\left(1-\frac{\langle \Delta \theta_\perp^2(\tau)\rangle}{\langle\Delta\theta^2(\infty)\rangle}\right)}\right]^{\frac{1}{0.95}},
\end{equation}
where \(\langle\Delta\theta^2(\infty)\rangle=\pi^2/24\). \Cref{MSAD_normalA} shows that that \(\langle \Delta \theta_\perp^2(\tau)\rangle_l\) closely follows original MSAD, \(\langle\Delta\theta_{xz}^2(\tau)\rangle+\langle\Delta\theta_{yz}^2(\tau)\rangle\).

\section{Approximation of out of plane fluctuation} 
 
Considering that the size of the nanorods are orders of magnitude smaller than the size of the GUVs we assume the membrane shape is nearly flat. Therefore, the shape of the membrane can be described by the height of the surface, \(z=h(x,y)\). The membrane normal then is determined by the gradient of the surface function, \(n=\nabla\Phi\), where \(\Phi\) has a form of \(\Phi=z-h(x,y)\). The local out of plane angle of the membrane is \(cos^2\theta=1/(n_x^2+n_y^2+1)\). Since \(n_x=\partial h/\partial x\) and \(n_y=\partial h/\partial y\), we can write \(tan^2\theta=(\partial h/\partial x)^2+(\partial h/\partial y)^2\), and considering paraxial approximation, \(\theta^2\approx (\partial h/\partial x)^2+(\partial h/\partial y)^2\). In a Fourier space, one can get \(\theta_q^2=q_x^2h_{xq}^2+q_y^2h_{yq}^2\), where \(q_{x,y}\) are the wave numbers in x and y directions. Finally applying binomial approximation, we estimate out of plane angle of the membrane based on the its local height in a Fourier space, \(\theta_q\approx qh_q\). 

Often, especially in small wave numbers, a lipid bilayer system is treated as a 2D incompressible sheet with small surface tension and bending modulus which exclusively govern its thermal curvature fluctuation. In this picture, due to Helfrich \cite{helfrich_elastic_1973}, a membrane can undergo shape change with no longitudinal deformation. Nevertheless, for a membrane with finite thickness, a local curvature change in the length scale of membrane thickness can generate longitudinal deformation in each monolayer. To develop a model for out of plane motion of GNR on the membrane, we start with Seifert-Langer model, which is based on hybrid curvature-dilational modes considering dynamics controlled by both bulk fluid dissipation and friction between monolayers \cite{seifert_viscous_1993}:
\begin{equation}\label{eq:seifert}
    \Big\langle h_q(t+\tau)h_q^*(t)\Big\rangle_t = \frac{k_BT}{\kappa_c q^4+\sigma q^2}\Big[A_1(q)e^{-\omega_1(q)\tau}+A_2(q)e^{-\omega_2(q) \tau}\Big],
\end{equation}
where \(A_2=1-A_1\).
Although determining amplitudes, \(A_{1,2}\), and relaxation rates, \(\omega_{1,2}\) theoretically stills remains complicated and not useful for fitting to the experimental data \cite{mell_fluctuation_2015}, in specific regimes the relative amplitudes of each relaxation mode can be estimated. These regimes are separated by the crossover waver numbers, \(q_1 \equiv 2\eta\varepsilon/b\tilde{\kappa}_c\) and \(q_2 \equiv \sqrt{2b/\eta_m}\). \(q_1\) indicates the crossover wave number separating hydrodynamic bending mode from inter-monolayer slipping mode (\(q_1\) is identified as \(q_c\) in the main text). \(q_2\) is the wave number in which membrane viscosity becomes the main dissipative mechanism rather than inter-monolayer friction. In the regime where \(q_1 < q < q_2\), \(\omega_2=\tilde{\kappa}_cq^3/4\eta\) becomes the relaxation rate for the bending mode with effective bending modulus of \(\tilde{\kappa}_c\), and \(\omega_1=\varepsilon\kappa_c q^2/2\tilde{\kappa_c}b\) becomes the relaxation rate for slipping mode with amplitudes of \(A_1=2\varepsilon d^2/\tilde{\kappa_c}=1-\kappa_/\tilde{\kappa}_c\). Hence, \cref{eq:seifert} in this regime can be simplified as
\begin{equation}\label{eq:simple}
    \Big\langle h_q(t+\tau)h_{-q}(t)\Big\rangle_t = \frac{k_BT}{\kappa_c q^4+\sigma q^2}\Big[(1-\frac{\kappa_c}{\tilde{\kappa}_c})e^{-\omega_1(q)\tau}+\frac{\kappa_c}{\tilde{\kappa}_c}e^{-\omega_2(q) \tau}\Big].
\end{equation}
And in this regime the approximation of \(\theta_q=qh_q\) yields
\begin{equation}\label{eq:corrangle1}
    \Big\langle \theta_q(t+\tau)\theta_{-q}(t)\Big\rangle_t = \frac{k_BT}{\kappa_c q^2+\sigma}\Big[(1-\frac{\kappa_c}{\tilde{\kappa}_c})e^{-\omega_1(q)\tau}+\frac{\kappa_c}{\tilde{\kappa}_c}e^{-\omega_2(q) \tau}\Big].
\end{equation}
The mean square displacement of the out of plane angle then can be calculated from the angle correlation function by summing over the effective wave numbers
\begin{dmath}\label{eq:msdn}
    \Delta \theta^2(\tau)=\big\langle \theta(t+\tau)-\theta(t)\big\rangle=
    \frac{1}{4\pi^2}\int_{q_{min}}^{q_{max}}\Big[ 2\langle \theta_q(t)\theta_{-q}(t) \rangle-2\langle \theta_q(t+\tau)\theta_{-q}(t) \rangle  \Big]2\pi qdq.
\end{dmath}
Where we considered \(\langle \theta_q(t)\theta_{-q}(t) \rangle\) to be constant over time as ergodicity requires. Using \cref{eq:simple} and \(\langle \theta_q\theta_{-q} \rangle=k_BT/(\kappa_c q^2+\sigma)\), we immediately obtain
\begin{dmath}
\label{MSAD_normal_eq}
    \langle\Delta\theta^2(\tau)\rangle=\frac{k_BT}{\pi}\int_{q_{min}}^{q_{max}}\frac{qdq}{\kappa_cq^2+\sigma}
    \left[1-\left(1-\frac{\kappa_c}{\tilde{\kappa}_c}\right)e^{-\omega_1\tau}-\frac{\kappa_c}{\tilde{\kappa}_c}e^{-\omega_2\tau}\right].
\end{dmath}
The integral is computed numerically using global adaptive quadrature method. 

\begin{figure}
    \centering
    \includegraphics[width=.5\textwidth]{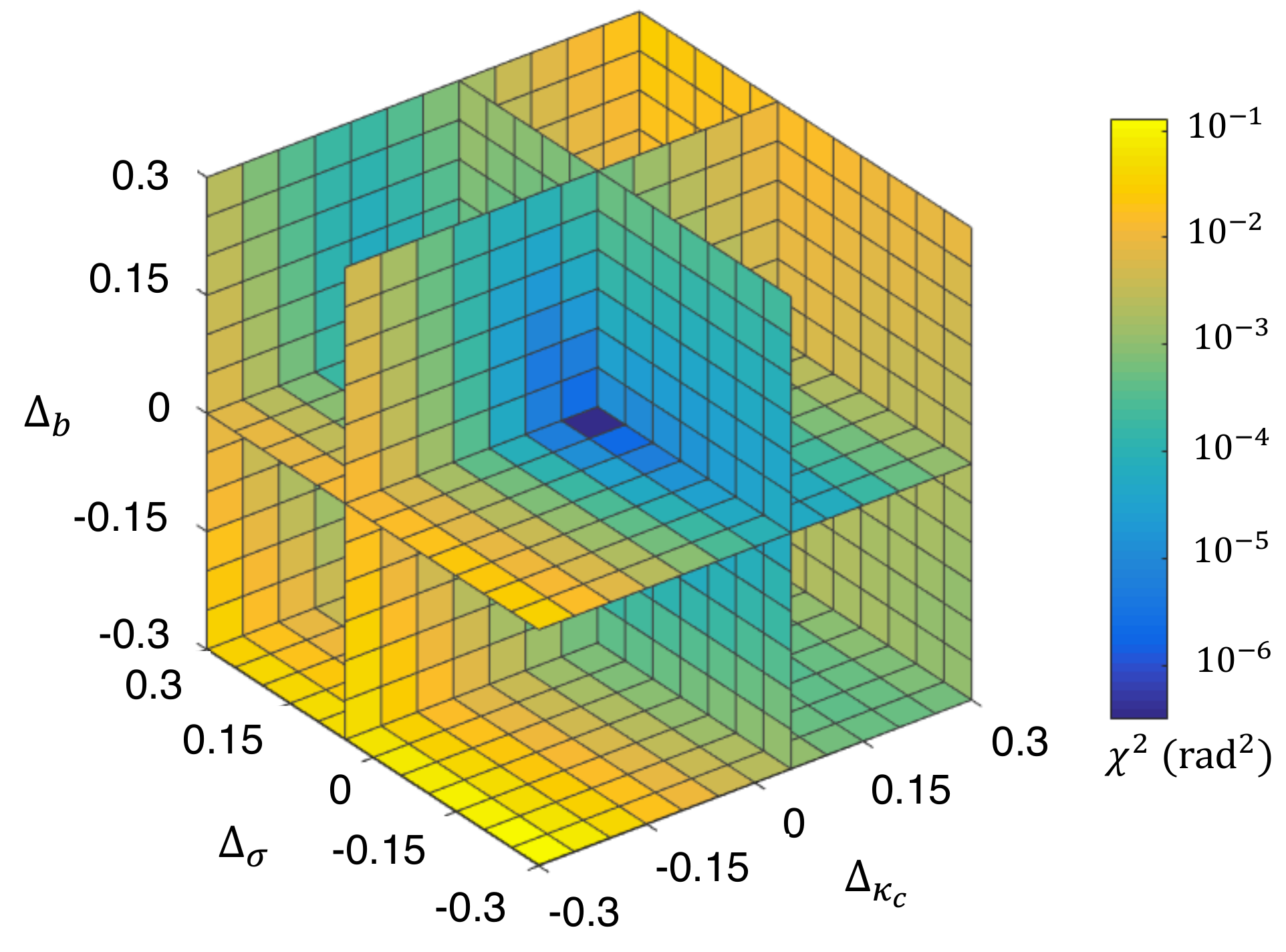}
    \caption{\(\chi^2\) test to evaluate sensitivity of the fitting for out of plane motion of GNR on the isoosmolar GUV with no cholesterol. \(\chi^2\) is calculated using Eqn. \ref{MSAD_normal_eq} and Eqn. \ref{eq:chi2} at \(\kappa=\kappa_{fit}+\Delta\kappa\), \(b=b_{fit}+\Delta b\), and \(\sigma=\sigma_{fit}+\Delta\sigma\), where subscription 'fit' indicates the values obtained from the fitting.}
    \label{fig:chi2}
\end{figure}

To evaluate sensitivity of the measurement to the fitting parameters we compute \(\Delta\theta^2(\tau)\) for sets of \(\kappa_c\), \(b\), and \(\sigma\) at logarithmically distributed lag-times using Eqn. \ref{MSAD_normal_eq}. \(\chi^2\) values is then calculated by comparison of measured and computed \(\Delta\theta^2(\tau)\):
\begin{dmath}\label{eq:chi2}
    \chi^2(\kappa,~b,~\sigma)=\sum_\tau{\frac{\big[\Delta\theta^2(\tau)_{exp}-\Delta\theta^2(\tau)|_{\kappa,b,\sigma}\big]^2}{\Delta\theta^2(\tau)|_{\kappa,b,\sigma}}}
\end{dmath}. 
\Cref{fig:chi2} shows that the \(\chi^2\) is the smallest at the values obtained from the fitting and increases rapidly as \(\kappa_c\), \(b\), and \(\sigma\) deviate from the fitted values. 

Furthermore, we evaluate the covariance of the out of plane angle of the nanorod, \(A(\tau)=\langle \theta_\perp(t+\tau)\theta_\perp (t) \rangle_t\), when it is only effected by the narrow range of undulation modes with wave number of \(q'-\delta q'/2<q<q'+\delta q'/2\). We start by integrating Eqn. \ref{eq:corrangle1} over the narrow range  of \(q\),
\begin{equation}
    A(\tau) = \frac{k_BT}{2\pi}\int_{q'-\delta q'/2}^{q'+\delta q'/2}{\frac{qdq}{\kappa_c q^2+\sigma}\Big[(1-\frac{\kappa_c}{\tilde{\kappa}_c})e^{-\omega_1(q)\tau}+\frac{\kappa_c}{\tilde{\kappa}_c}e^{-\omega_2(q) \tau}\Big]},
\end{equation}
and simply use trapezoidal rule to approximate this integral
\begin{equation}
    A(\tau) = \frac{k_BT}{2\pi}\frac{q'\delta q'}{\kappa_c q^{'2}+\sigma}\Big[(1-\frac{\kappa_c}{\tilde{\kappa}_c})e^{-\omega_1(q')\tau}+\frac{\kappa_c}{\tilde{\kappa}_c}e^{-\omega_2(q') \tau}\Big],
\end{equation}
Since the narrow range of q includes the wave number set by the size of the GNR, we expect that \(\sigma\ll\kappa_c q^{'2}\) and \(\kappa_c\ll\tilde{\kappa}_c\) which simplifies above equation as
\begin{equation}
    A(\tau) = \frac{k_BT}{2\pi \kappa_c}\frac{\delta q}{q}e^{-\omega_1(q')\tau}.
\end{equation}

We assume that the lower range of the undulation wave number is set by the typical distance between pinning sites of the membrane to the cell cortex. While this value varies in different cells it is expected to be close to the typical value of the F-actin mesh size of the cell cortex, \(l_{pin}\approx \SI{200}{\nano\meter}\) \cite{morone_three-dimensional_2006}. The upper range of the undulation wave number is set by the size of the GNR. Hence, we estimate that \(\delta q/q=2(q_{max}-q_{min})/(q_{max}+q_{min})\approx 2(l_{pin}-l)/(l_{pin}+l)\). 

\section{Membrane-GNR  Monte Carlo model}
 
{\it{Dynamically triangulated Monte Carlo  model for membrane:}}~We consider a patch of  fluctuating membrane bounded by a square frame of size $L$ with periodic boundary conditions in lateral directions. The continuum  surface of the membrane is discretized into $N_{t}$ triangles which connects  $N_v$ vertex points through $N_l$ links~\cite{ho_self-avoiding_1989}. The membrane conformations are governed by elastic energy of the surface given by Canham-Helfrich approximation for  bilayer membranes~\cite{helfrich_elastic_1973}; the discretized form of hamiltonian is given by 

\begin{equation}
\mathbf{H}_{\mathrm{elastic}}=\frac{\kappa_c}{2}{\sum_{v=1}^{N_v}(c_{1,v}+ c_{2,v})^2 }A_v.
\label{sm_eqn1}
\end{equation}
Where, $\kappa_c$ is the bending rigidity of membrane, $A_v$ is the area and $c_{1,v}$ and $c_{2,v}$ are the principal curvatures at vertex $v$.  To compute the principal  curvatures  we follow the method introduced by Ramakrishnan et al. ~\cite{ramakrishnan_monte_2010}.  The self avoidance of surface is satisfied by setting the link length $l$ as $a_0 \le l \le \sqrt{3}a_0$ where $a_0$ is the hard sphere radius of the vertices. 

In DTMC the membrane surface is equilibrated through two independent Monte carlo moves namely vertex moves and link flips. In  a vertex move the position $X_v$ of a vertex is updated to a new random position $X_v+\delta X_p $ within a cubic box of size $\epsilon_v$. The size of the box is selected such that $50\%$ of moves are accepted. This move simulates the thermal fluctuations  and allows the membrane shape to relax to an equilibrium conformation. In a link flip move: a randomly selected link, between two triangles is disconnected, and new a link is established with the unconnected vertices of the same two triangles. This move makes the triangulation dynamic and preserves the fluid nature of the bilayer membranes by ensuring the in-plane displacement of vertices. MC moves are accepted through the Metropolis algorithm.

\begin{figure}
\centering
\includegraphics[scale=0.5]{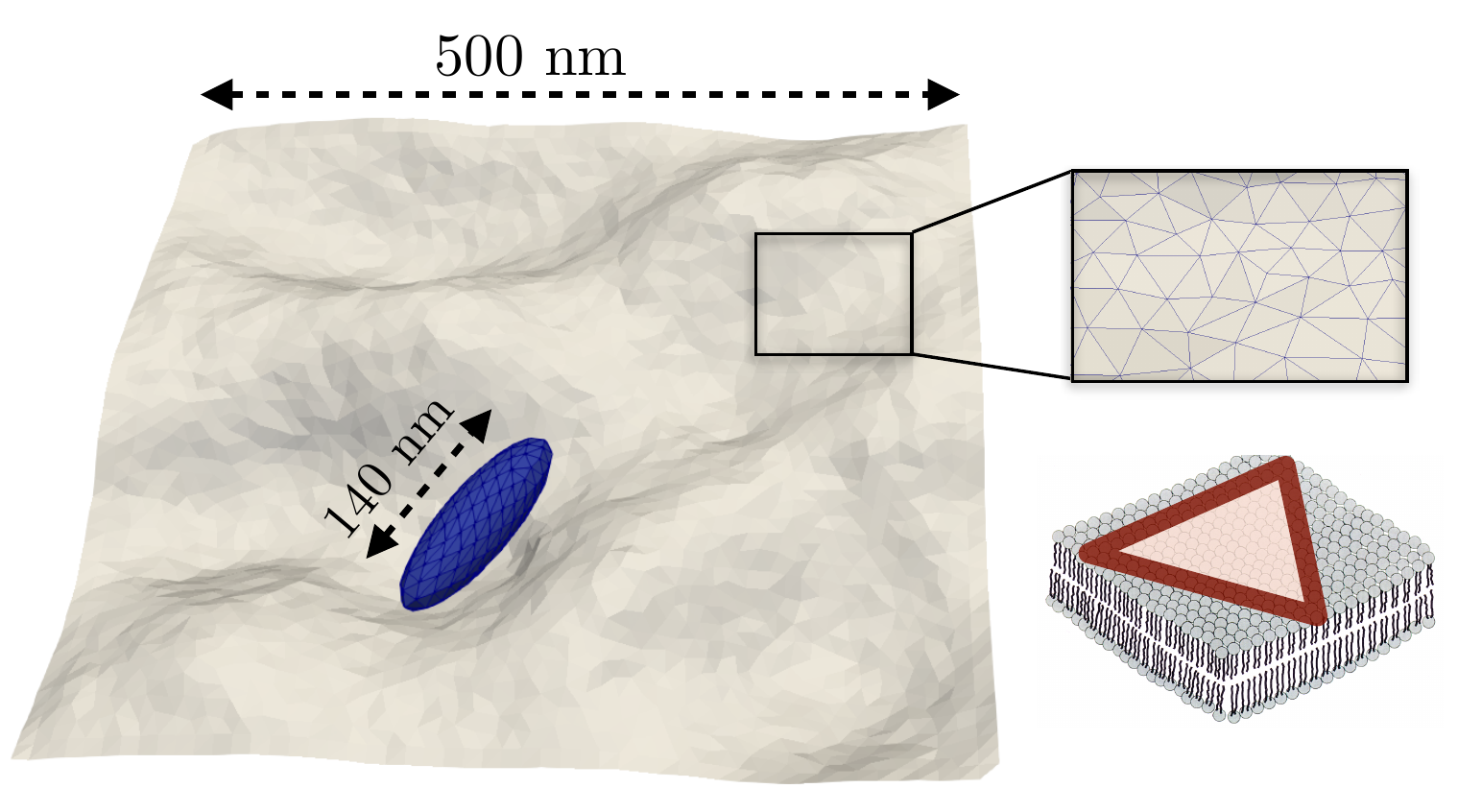}
\caption{\label{fig_SM1} A snapshot of triangulated surface membrane with a bound GNR.}
\end{figure}

{\it{Ellipsoidal model for GNR:}}~  In our representation the GNR is modeled as an ellipsoid with three principal dimensions: \(a=\SI{70}{\nano\meter}\), \(b=c=\SI{20}{\nano\meter}\). The surface of the ellipsoid is discretized into 162  equally spaced points. The discrete points on the model GNR surface interact with membrane vertices via a truncated Lennard Jones potential given as 

\begin{figure}
\centering
\includegraphics[scale=0.5]{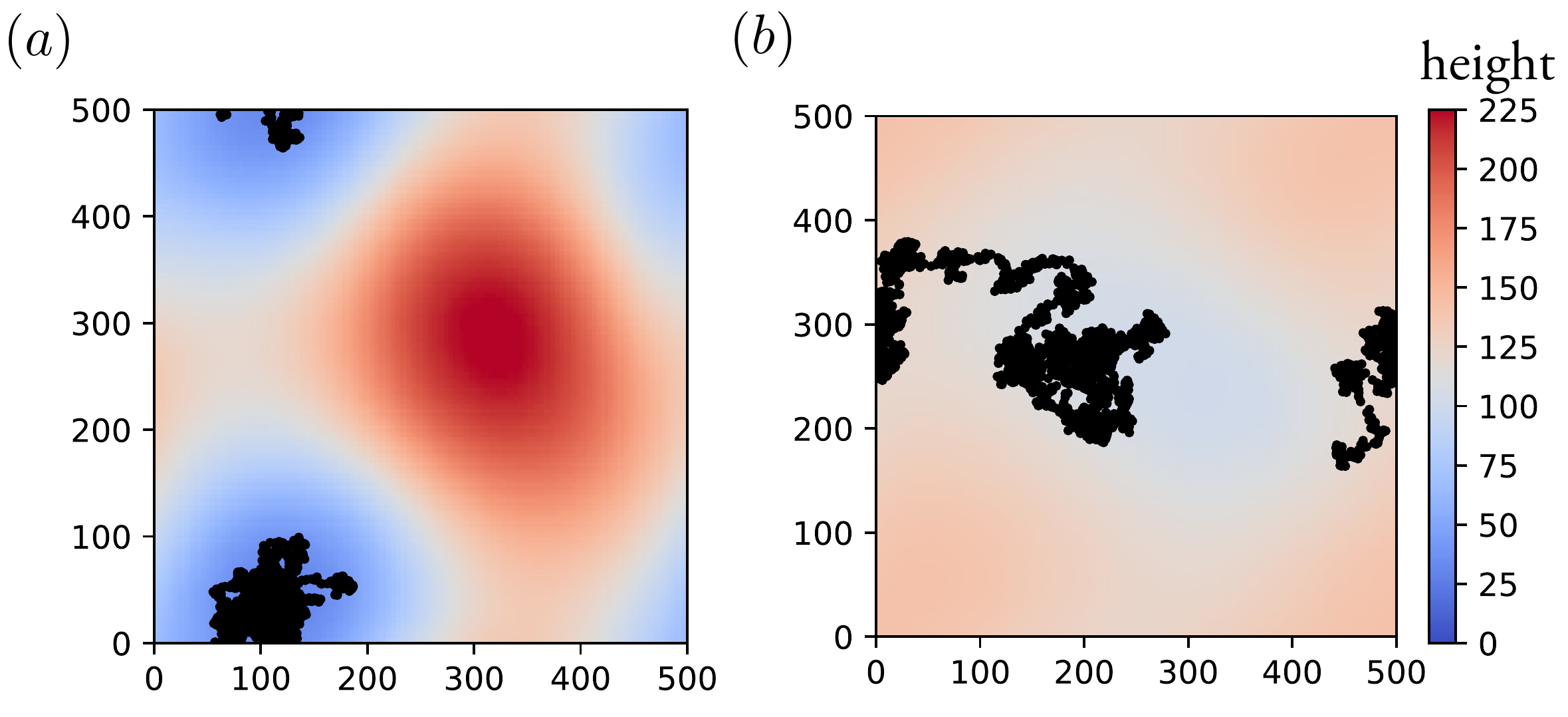}
\caption{\label{fig_SM2} Trajectory of model GNR on membrane surface at (a) high  and (b) low  $A_{\mathrm{ex}}$ for a time span of $2.5 \times 10^6$ MC steps.}
\end{figure}

\begin{equation}
V_{LJ}(r) = 4\epsilon_{LJ} \left[ \left(\frac{\sigma}{r}\right)^{12} - \left(\frac{\sigma}{r}\right)^6 \right]\begin{cases} 
1  \text{\vspace{1cm} if }r\leq r_t,\\
S(r)\text{\vspace{1cm} if }r_t\leq r \leq r_c,
\end{cases}
\label{}
\end{equation}
where $r$ is the distance of the dimer spheres from the membrane vertices  and  $S(r)$ is a function to smoothly truncate the potential, given as 
\begin{equation}
S(r)=1-\frac{(r-r_t)^2 (3r_c-2r-r_t)}{(r_c-r_t)^3}.
\end{equation}
\begin{figure}
\centering
\includegraphics[scale=0.5]{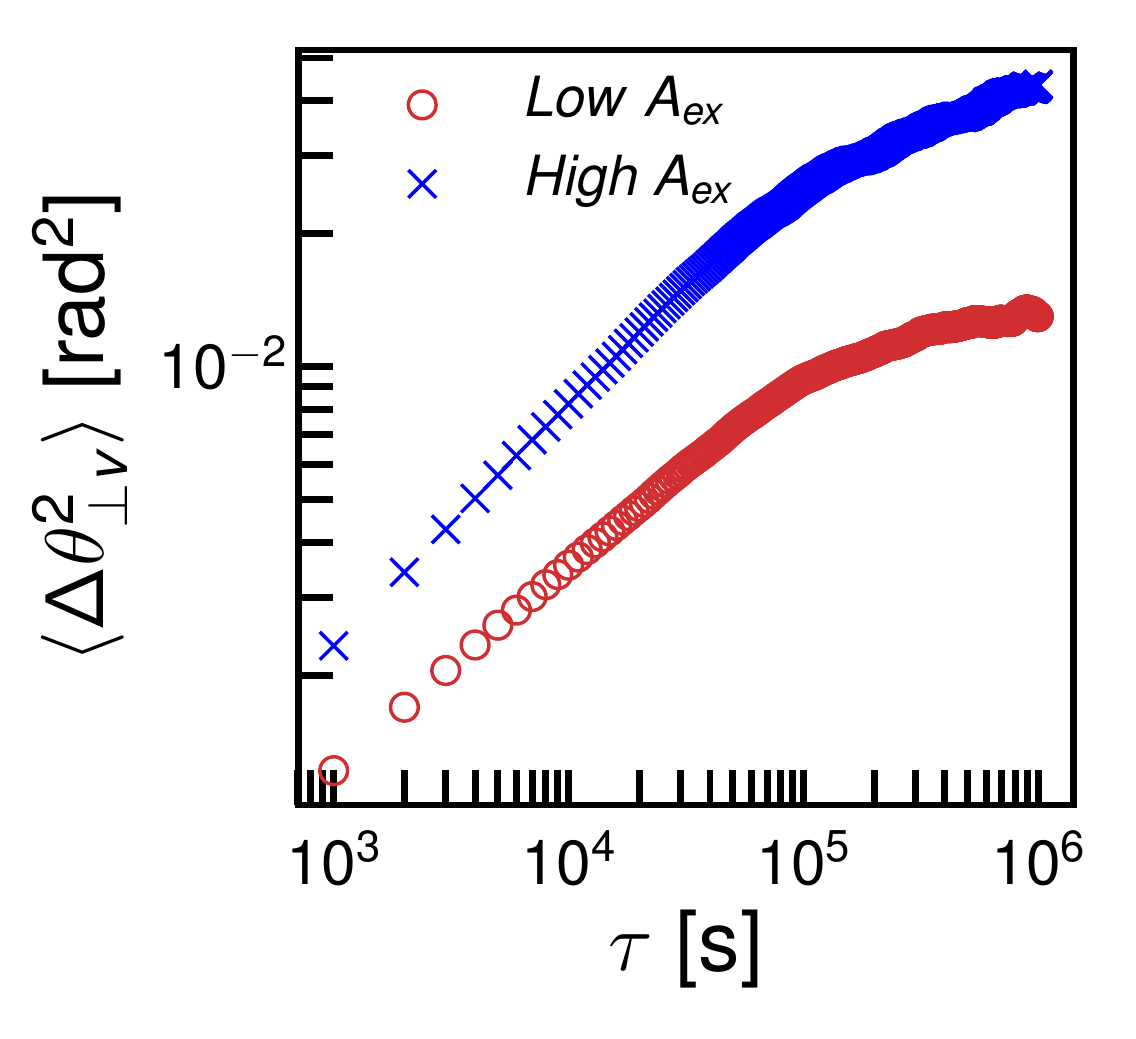}
\caption{\label{fig_SM3} Mean squared angular deviation of $\Delta \theta_{\bot v}$ where  $\Delta \theta_{\bot v}=\cos^{-1}\left(\frac{1}{n}\sum_v \hat{n_{t}^v}.\hat{z}\right)-\cos^{-1}\left(\frac{1}{n}\sum_v\hat{n_{0}^v}.\hat{z}\right)$, n runs over all membrane vertices that are attached to GNR and $\hat{n}_{t}^v$ is the membrane vertex normal.}
\end{figure}

We choose $r_t=1.1688\sigma$ and $r_c=1.3636\sigma $. We choose \(\sigma=\SI{15}{\nano\meter}\) and $\epsilon_{LJ}=2.5~k_BT$. A snapshot of GNR-membrane model is given in \cref{fig_SM1}.
\begin{figure}
\includegraphics[scale=0.4]{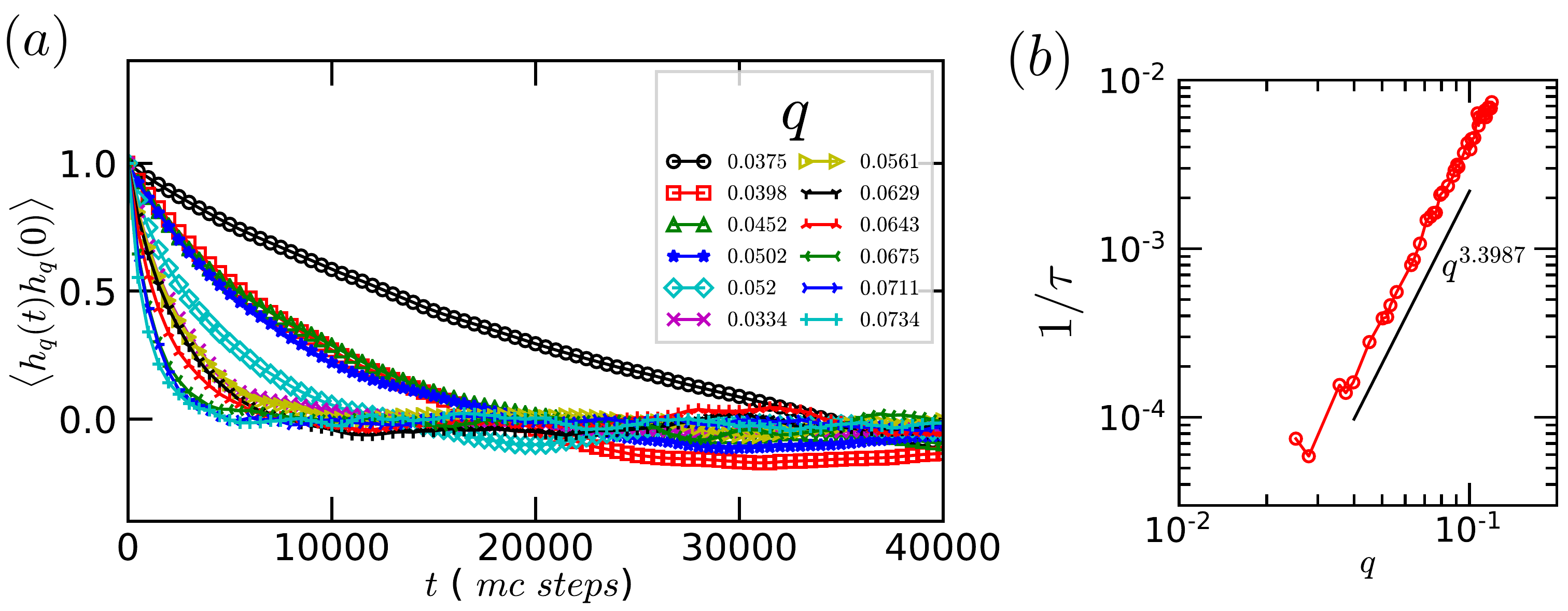}
\caption{\label{fig_SM4} (a) Membrane height correlation in frequency space. (b) Relaxation rate as a function of $q$ obtained by fitting the correlation curve in (a) to exponential.}
\end{figure}

\begin{center}
\begin{table}[htbp]
\begin{tabular}[t]{c c  c c}
\hline
$A_{ex} (\%)$ \ \ \ & \ \  \ $\sigma (k_BT/nm^2)$ \ \ \ & \ \ \ $D$($nm^2$/mcs steps) \\
\hline
  2.3  \ \ \ &   \ \ \    0.0216 \ \ \ & \ \ \ 0.0069 \\
  3.3  \ \ \ &   \ \ \    0.0191\ \ \ & \ \ \ 0.0058 \\
  4.5 \ \ \ & \ \ \       0.0173\ \ \ & \ \ \ 0.0055 \\
  16.1 \ \ \ & \ \ \     0.0172\ \ \ & \ \ \ 0.0029\\
  20.0 \ \ \ & \ \ \     0.0166 \ \ \ & \ \ \ 0.0031\\
  21.3 \ \ \ & \ \ \  0.0153\ \ \ & \ \ \ 0.0031 \\
  \hline
	\hline
\end{tabular}
\caption{Membrane tension and ellipsoidal nanoparticle diffusion coefficient as a function of $A_{ex}$.}
\label{I}
\end{table}
\end{center}

\section{GNR motion on a plasma membrane}
\begin{figure}
    \centering
    \includegraphics[width=.35\textwidth]{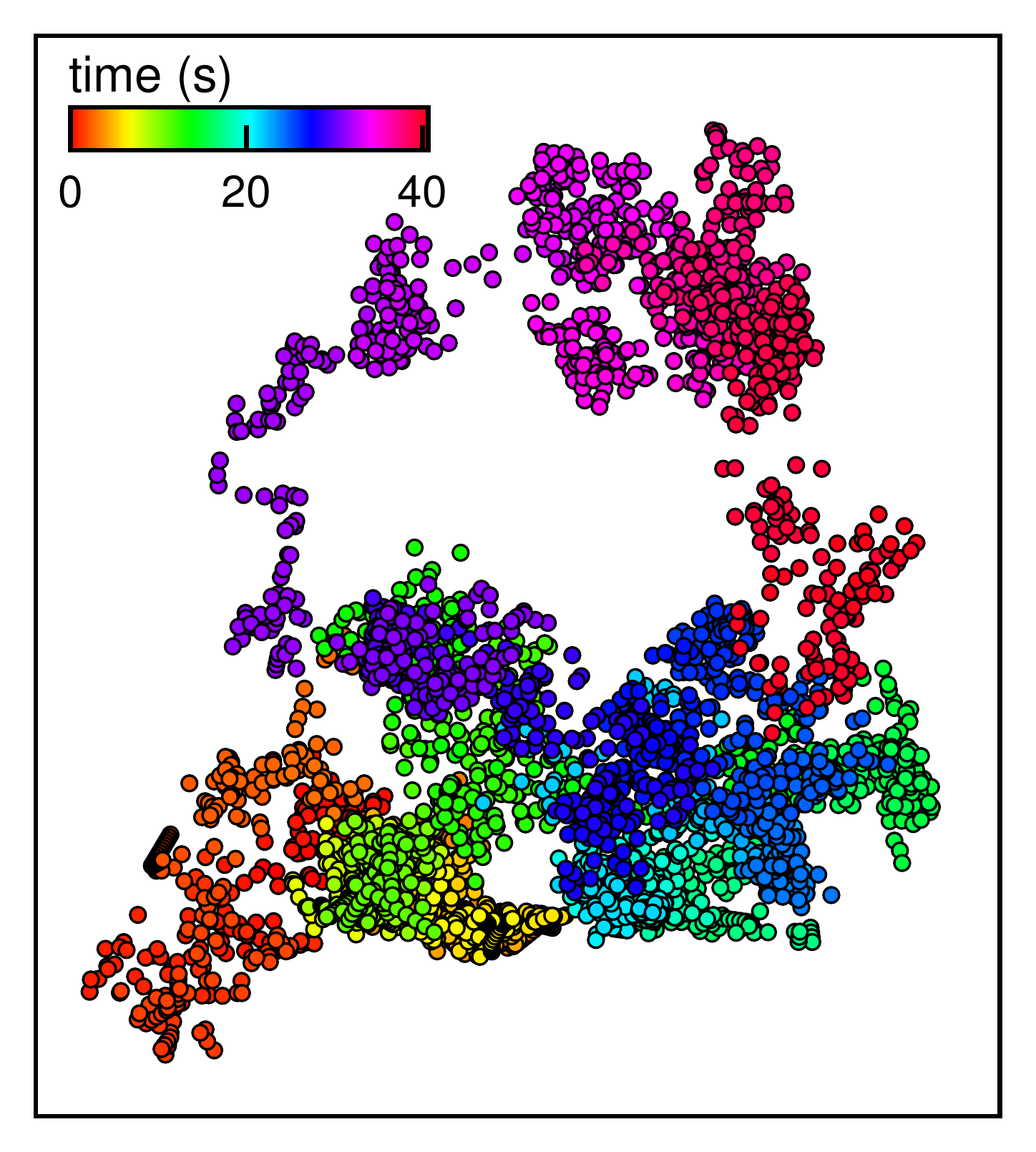}
    \caption{Trajectory of the GNR diffusing randomly on the plasma membrane of a Huh7 cell.}
    \label{traj_long}
\end{figure}

\begin{figure}
    \centering
    \includegraphics[width=.5\textwidth]{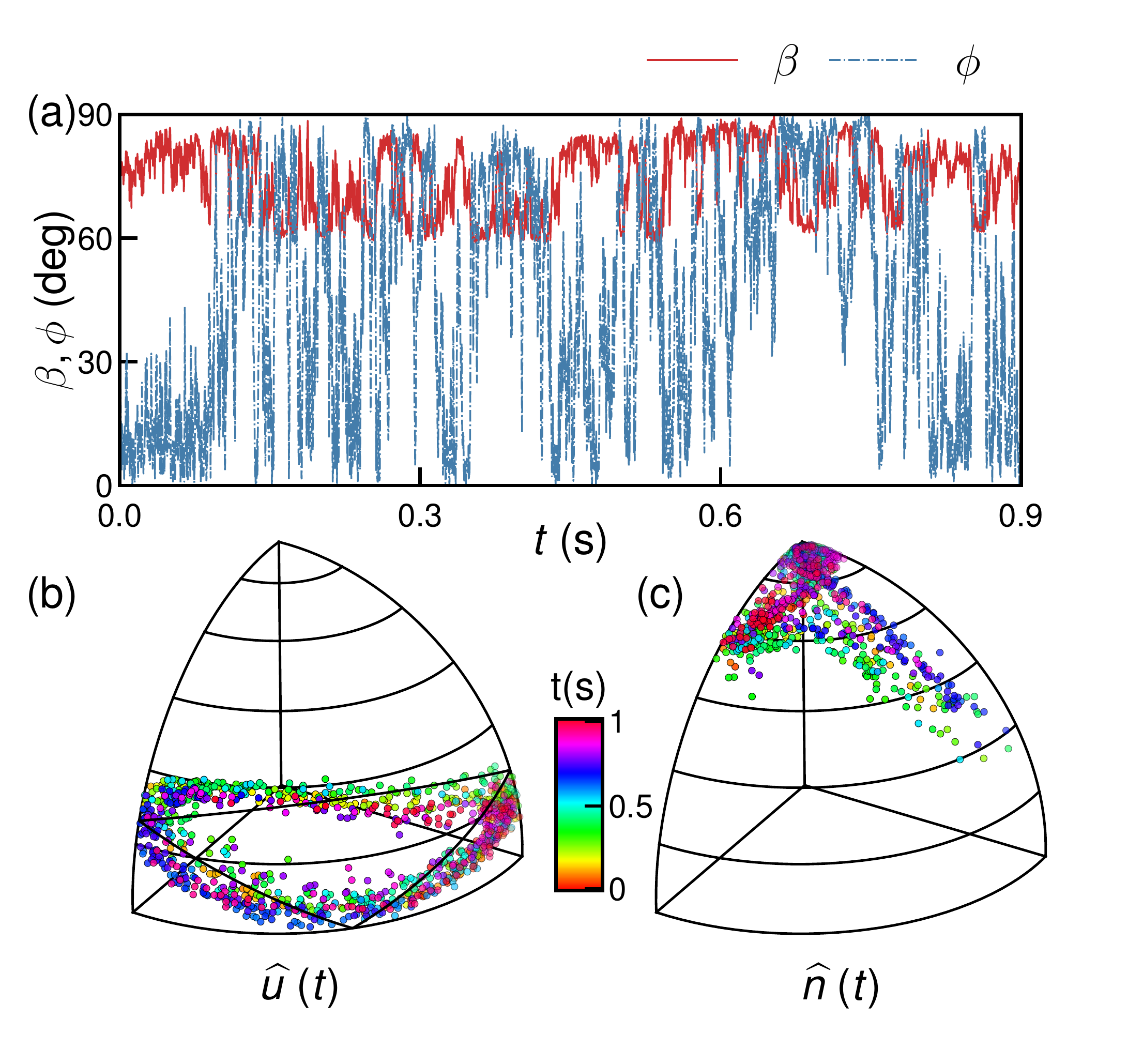}
    \caption{Motion of a GNR on a plasma membrane of a Huh7 cell. a) Azimuthal, \(\phi\), and polar, \(\beta\), angles. b) GNR orientation, \(\hat{\vec{u}}\), indicates that it is limited to 2D plane which slowly fluctuates. c) The normal vector of the plasma membrane, \(\hat{\vec{n}}\) estimated from \(\hat{\vec{u}}\).}
    \label{traj_angle}
\end{figure}

A 2D trajectory of the GNR moving on the plasma membrane is shown in \cref{traj_long}. The GNR performs 2D random walk on the membrane and stays in focus for more than 5 minutes. 
The azimuthal and polar angles of the nanorod and its orientation vector, \(\hat{\vec{u}}\), and the reconstructed normal vector of the plasma membrane, \(\hat{\vec{n}}\),  are shown in \cref{traj_angle}.

\section{Videos}

\textbf{Video S1:} Random rotational motion of a rod on a tilted plane.\\
\textbf{Video S2:} The simulated motion of a nanorod on simulated membranes with two different values of excess membrane area, corresponding to the tense and floppy membranes.\\

\bibliography{Ref_GUV.bib}